\documentclass[final, number, 3p, 12pt]{elsarticle}
\biboptions{sort&compress}

\usepackage{epsfig}

\usepackage{amssymb}
\usepackage{amsmath}
\usepackage{hyperref}
\usepackage{float}
\usepackage[normalem]{ulem}

\usepackage{svg}

\newcommand{\fig}[1]{\figurename~\ref{#1}}
\newcommand{\tbl}[1]{\tablename~\ref{#1}}

\begin{document}

\begin{frontmatter}



\title{A Robust All-Mach Six-Equation Diffuse-Interface Method for Multiphase Flows with Surface Tension}


           
\author[label1]{Ghanshyam Bharate}
\author[label1]{J.C. Mandal}
\affiliation[label1]{organization={Department of Aerospace Engineering, Indian Institute of Technology Bombay},  city={Mumbai}, postcode={400076}, country={India}}

\begin{abstract}
A robust finite-volume framework is presented for the simulation of compressible multiphase flows with surface tension across a wide range of Mach numbers. The method is based on a two-pressure, six-equation diffuse interface model incorporating viscous, gravitational, and capillary effects through the continuum surface force formulation. To consistently account for capillary-induced pressure jumps, an HLLC Riemann solver is developed using generalized Riemann invariant analysis, and the instantaneous pressure relaxation procedure is modified to preserve the Laplace pressure jump during phase equilibration. To overcome the excessive numerical diffusion of conventional approximate Riemann solvers in the low-Mach regime, a robust low-Mach correction is proposed by extending our previous formulation with a modified scaling strategy that remains stable in regions of strong pressure variation. The resulting method retains accuracy from nearly incompressible flows to compressible regimes while preserving the robustness of the six-equation formulation. The numerical framework is validated using a series of benchmark problems involving surface tension, viscosity, gravity, and compressibility. The results demonstrate accurate prediction of interface dynamics, capillary pressure, and low-Mach flow features, while significantly reducing numerical dissipation without compromising stability. The proposed methodology provides an efficient and reliable approach for the simulation of complex multiphase flows spanning a broad range of flow regimes.
\end{abstract}



\end{frontmatter}


\section{Introduction}

Numerical simulation of compressible multiphase flows is of fundamental importance in numerous engineering and scientific applications, including cavitation, droplet dynamics, bubble collapse, atomization, shock–bubble interactions, additive manufacturing, and microfluidics. These problems often involve large density and pressure gradients, moving interfaces, surface tension, viscous effects, and, in many practical situations, flow speeds ranging from nearly incompressible to highly compressible regimes. Developing numerical methods that remain accurate and robust over such a broad range of physical conditions continues to be a significant challenge.

Among the available approaches for interface-resolved multiphase computations, diffuse interface methods (DIMs) have emerged as an attractive alternative to sharp-interface techniques because they naturally accommodate large interface deformations, topological changes, and compressibility within a conservative finite-volume framework. Since the pioneering work of \citet{saurel1999}, diffuse interface methods have been successfully applied to a variety of compressible multiphase flow problems. Their principal advantage lies in replacing the sharp material interface with a narrow transition region, thereby avoiding explicit interface reconstruction. However, the inherent numerical diffusion associated with this formulation causes interface smearing, which may degrade the accuracy of interfacial dynamics. Considerable effort has therefore been devoted to reducing numerical diffusion through interface-compression techniques \cite{shukla2010interface, fiveeq_st_csf_2017,tiwari2013diffuse}, THINC (Tangent of Hyperbola for INterface Capturing) reconstruction scheme \cite{nonomura2014simple, shyue2014eulerian, garrick2017interface}, and compressive limiting strategies within the MUSCL framework \citet{overbee}. These developments have substantially improved the ability of diffuse interface methods to capture complex interface evolution while preserving numerical robustness.

Several mathematical models have been proposed within the diffuse interface framework. The five-equation model \cite{allaire2002five, massoni2002proposition}, has become one of the most widely used formulations for compressible multiphase flows because of its simplicity and computational efficiency. An alternative formulation, commonly referred to as the five-equation reduced model \cite{kapila2001, murrone2005five}, is obtained from the non-equilibrium seven-equation model under the assumption of instantaneous pressure and velocity relaxation. Compared with the original five-equation model, the reduced formulation includes an additional compressibility correction term in the volume-fraction equation, enabling accurate prediction of important phenomena such as Rayleigh bubble collapse \cite{tiwari2013diffuse,rasthofer2017large}, while satisfying the second law of thermodynamics \cite{murrone2005five}.

Despite these attractive properties, the five-equation reduced model suffers from several numerical deficiencies, including difficulties in maintaining volume-fraction positivity under strong expansion or compression \cite{beig2018temperatures} and the occurrence of a non-monotonic mixture sound speed, which may introduce spurious sonic points inside diffuse interfaces  \cite{Saurel2009SimpleAE,petitpas2007relaxation}. To overcome these issues, pressure non-equilibrium is retained by employing the two-pressure six-equation model \cite{Saurel2009SimpleAE}. Combined with an instantaneous pressure-relaxation procedure, this formulation recovers the equilibrium solution while avoiding many of the numerical difficulties encountered in the reduced five-equation model. Consequently, the six-equation model has become a preferred framework for compressible diffuse interface simulations \cite{Saurel2009SimpleAE,pelanti2014mixture,yu2023numerical,sixeq_st_css_2017,schmidmayer2020assessment,hong2022numerical} and is adopted in the present work.

Although diffuse interface methods have achieved considerable success for compressible flows, they generally exhibit excessive numerical dissipation at low Mach numbers. This behaviour originates from approximate Riemann solvers whose artificial dissipation scales with the acoustic wave speed rather than the fluid velocity. Various preconditioning techniques \cite{murrone2008,braconnier2009all,liquidgas2013,pelanti2017low,guillard1999behaviour,guillard2004behavior,pelanti2018wave} have been proposed to alleviate this deficiency and have been successfully extended to multiphase flows. However, preconditioning methods suffer from well-known drawbacks, including the global Mach-number cut-off problem and severe time-step restrictions in explicit time integration \cite{li2008all, birken2005stability}. In our previous work \cite{bharate2025enhanced}, we proposed an alternative low-Mach correction based on modifying the velocity jumps within the Riemann solver. The approach is straightforward to implement, avoids additional time-step constraints, and substantially improves accuracy for low-Mach multiphase flows. Nevertheless, its performance has not been investigated for flows involving surface tension and viscosity, and the original scaling strategy may become unstable in regions with strong pressure gradients.

Accurate representation of surface tension constitutes another essential requirement for many multiphase applications. Several numerical formulations have been developed to incorporate capillary effects into diffuse interface methods \cite{fiveeq_st_css_2005,fiveeq_st_csf_2017,seveneq_st_path_2015,sixeq_st_css_2017}. Exact \cite{fiveeq_st_css_2005} and HLLC-type approximate \cite{fiveeq_st_csf_2017} Riemann solvers have been proposed for five-equation models by accounting for the Laplace pressure jump across material interfaces. \citet{seveneq_st_path_2015} developed a path-conservative Osher-type method for the seven-equation model including capillary effects, whereas \citet{sixeq_st_css_2017} introduced a three-step operator-splitting procedure for the six-equation model based on a curl-free formulation of the surface-tension force. Although these methods have demonstrated promising performance, existing six-equation formulations do not explicitly modify either the Riemann solver or the pressure-relaxation procedure to consistently preserve the capillary pressure jump. Consequently, a formulation that directly incorporates capillary effects within the wave propagation and relaxation processes remains unavailable.

The objective of the present work is to develop a robust numerical framework for the simulation of multiphase flows with surface tension over a broad range of Mach numbers. The proposed method is based on the two-pressure six-equation diffuse interface model with viscous, gravitational, and capillary effects represented through the continuum surface force formulation. An HLLC Riemann solver incorporating capillary pressure jumps is derived using generalized Riemann invariant analysis, and the instantaneous pressure-relaxation procedure is modified to preserve the Laplace pressure jump during phase equilibration. Furthermore, a more robust low-Mach correction is proposed by extending our previous formulation with a modified scaling strategy that remains stable in regions of large pressure variation while retaining accuracy in the incompressible limit. The resulting numerical framework is assessed using a series of benchmark problems involving surface tension, viscosity, gravity, and compressibility, demonstrating its accuracy, robustness, and applicability over a wide range of multiphase flow regimes.

Finally, the remainder of the paper is organized as follows. Section \ref{sec:model}, presents the governing six-equation multiphase model. Section \ref{sec:Nmethod} describes the finite-volume discretization, including the HLLC Riemann solver, surface-tension treatment, curvature evaluation, viscous discretization, and pressure-relaxation procedure. Section \ref{sec:LowMach} introduces the modified low-Mach correction. Numerical validations are presented in Section \ref{sec:results}, followed by concluding remarks in Section \ref{sec:conclusion}.

\section{Multiphase model}\label{sec:model}
Six-equation multiphase model~\cite{kapila2001, Saurel2009SimpleAE, six_eq_viscous} including viscous and surface tension effects can be written as   
\begin{equation}\label{eq:model}
	\begin{split}
		&\frac{\partial \alpha_1}{\partial t} + \mathbf{u} \cdot \nabla \alpha_1 = K_p (p_1 - p_2), \\
		&\frac{\partial \alpha_1 \rho_1}{\partial t} + \nabla \cdot (\alpha_1 \rho_1 \mathbf{u}) = 0, \\
		&\frac{\partial \alpha_2 \rho_2}{\partial t} + \nabla \cdot (\alpha_2 \rho_2 \mathbf{u}) = 0, \\
		&\frac{\partial \rho \mathbf{u}}{\partial t} + \nabla \cdot (\rho \mathbf{u} \otimes \mathbf{u}) + \nabla p = \nabla \cdot \mathsf{\tau} + \mathbf{f}_{\sigma} + \rho \mathbf{g}, \\
		&\frac{\partial \alpha_1 \rho_1 e_1}{\partial t} + \nabla \cdot (\alpha_1 \rho_1 e_1 \mathbf{u}) + \alpha_1 p_1 \nabla \cdot \mathbf{u} =
		\alpha_1 \mathcal{\tau}_1 : \nabla \mathbf{u}  - K_p p_I (p_1 - p_2), \\
		&\frac{\partial \alpha_2 \rho_2 e_2}{\partial t} + \nabla \cdot (\alpha_2 \rho_2 e_2 \mathbf{u}) + \alpha_2 p_2 \nabla \cdot \mathbf{u} =
		\alpha_j \mathcal{\tau}_2 : \nabla \mathbf{u}  + K_p p_I (p_1 - p_2).
	\end{split}
\end{equation}
Here $\alpha_j, \rho_j, p_j, e_j$ are the volume fraction, density, pressure, and specific internal energy of the phase $j$ respectively. The velocity is represented by $\mathbf{u}$ and $\mathbf{g}$ is gravitational acceleration. The mixture density and mixture pressure is denoted by $\rho$ and $p$ respectively. The expressions for these mixture quantities are given by
\begin{equation}
	\rho = \sum_{j} \alpha_j \rho_j , \quad p = \sum_j \alpha_j p_j.
\end{equation} 
On the right side of the volume fraction and specific energy equation we have pressure relaxation terms, $ K_p (p_1 - p_2)$ and $p_I K_p (p_1 - p_2)$. Here $K_p$ and $p_I$ are relaxation rate and interface pressure.  Interface pressure $(p_I)$ appearing in the relaxation terms is defined as \cite{Saurel2009SimpleAE,pelanti2014mixture} 
\begin{equation}
	p_I = \frac{( \rho_1 a_1) p_2 + ( \rho_2 a_2) p_1}{\rho_1 a_1 + \rho_2 a_2}.
\end{equation}

The system of equations~\eqref{eq:model} is closed by equation of state. Here stiffened gas equation of state (SG EOS) is used for each phase. The expressions of the specific internal energy $(e_j)$ and speed of sound $(a_j)$ for SG EOS can be written as 
\begin{equation}\label{eq:EOS}
	e_j  = \frac{p_j + \gamma_j \pi_j}{\rho_j \left(\gamma_j -1 \right)}, \quad a_j = \sqrt{\frac{\gamma_j \left(p_j + \pi_j \right)}{\rho_j}}.
\end{equation}
There are two viscous tensor terms appearing in the system \eqref{eq:model}, one is the mixture stress $\mathsf{\tau}$ and another is phasic stress $\mathcal{\tau}_j$. The both stress terms are assumed to be equal $ \left( \mathsf{\tau}_j = \mathsf{\tau} \right)$ \cite{six_eq_viscous,allmach_pressure_2021}. The stress tensor is defined as 
\begin{equation}
	\mathsf{\tau}  = \left[ \begin{array}{cc}
		\tau_{xx} & \tau_{xy} \\
		\tau_{yx} & \tau_{yy}
	\end{array} \right] =  \mu \left[\begin{array}{cc} \frac{4}{3} \frac{\partial u}{\partial x} - \frac{2}{3}  \frac{\partial v}{\partial y} & \frac{\partial u}{\partial y} +  \frac{\partial v}{\partial x} \\ [5pt] \frac{\partial u}{\partial y} +  \frac{\partial v}{\partial x} &  \frac{4}{3} \frac{\partial v}{\partial y} - \frac{2}{3} \frac{\partial u}{\partial x} \end{array} \right].
\end{equation}
The mixture viscosity $(\mu)$ can be defined as simple average of the phasic viscosity \cite{six_eq_viscous, fiveeq_st_css_2005}.
\begin{equation}
	\mu = \sum_j \alpha_j \mu_j. 
\end{equation}
The surface tension effects are accounted using continuum surface force (CSF) approach \cite{brackbill1992continuum}, where the capillary forces are modeled as volumetric source term in the multiphase system. This method is followed by many authors \cite{braconnier2009all, fiveeq_st_csf_2017,fiveeq_st_csf_2021, le2014towards, panchal2023seven} in case of compressible multiphase models. Surface tension force $\mathbf{f}_{\sigma}$ in the multiphase system \eqref{eq:model} is defined as
\begin{equation}
	\mathbf{f}_{\sigma} = \sigma \kappa \nabla \alpha_1.
\end{equation}
Where, $\sigma$ is the surface tension coefficient and $\kappa$ is the curvature of interface.  

The multiphase model \eqref{eq:model} includes phasic internal energy equations. There is no gurantee that numerical discretization of the these internal energy equations will conserve the mixture total energy, which is essential in the presence of shock waves. To address this issue, \citet{Saurel2009SimpleAE} introduced an extra mixture total energy equation to ensure total energy conservation. The mixture total energy equation can be written as
\begin{equation}
	\frac{\partial \rho E}{\partial t} + \nabla \cdot (\rho E \mathbf{u} + p \mathbf{u})  = \nabla \cdot \left( \mathcal{\tau} \cdot \mathbf{u} \right)	+  \mathbf{u} \cdot \mathbf{f}_{\sigma} + \rho \mathbf{g} \cdot \mathbf{u}.
\end{equation}
The mixture total energy $\rho E$ is defined as 
\begin{equation}
	\rho E = \rho e + \frac{1}{2} \rho \left| \mathbf{u} \right|^2.
\end{equation}
Here, $\rho e$ is the mixture internal energy it is expressed as
\begin{equation}
	\rho e = \sum_{j} \alpha_j \rho_j e_j.
\end{equation}

\section{Numerical Method}\label{sec:Nmethod}
Following the methods described in the literature \cite{Saurel2009SimpleAE,overbee,schmidmayer2020assessment} and our previous work \cite{bharate2025enhanced}, the six-equation model is solved using two step procedure. The first step, referred to as the evolution step, solves the multiphase system \eqref{eq:model} in the absence of relaxation terms. In the second step, only the pressure relaxation terms are retained, forming a system of ordinary differential equations (ODEs). These steps are elaborated in Sections \ref{sec:evol_step} and \ref{sec:relax_step}.

 \begin{figure}[H]
	\centering
	\includegraphics[width=0.75\textwidth]{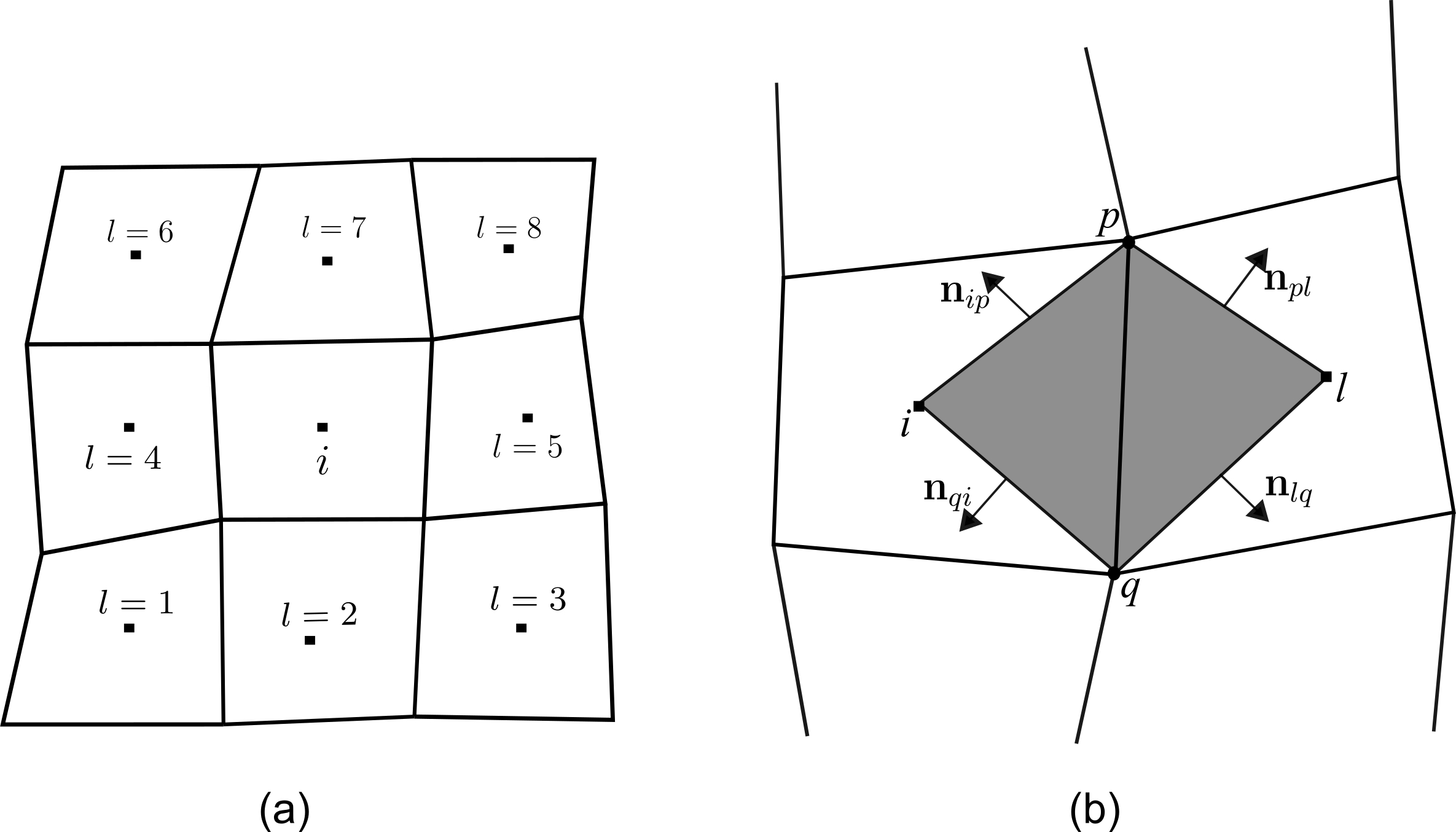}
	\caption{(a) Quadrilateral cell stencil for curvature estimation. (b) Estimation of velocity gradients for viscous fluxes at face $pq$ via the Green-Gauss method on a diamond-shaped control volume.}
	\label{fig:FVM1}
\end{figure}

\subsection{Evolution step}\label{sec:evol_step}
This section details the solution of system~\eqref{eq:model2} in the absence of relaxation effects. By omitting the pressure relaxation term, the multiphase system from Eq.~\eqref{eq:model} can be expressed in a compact form as: 
\begin{equation}\label{eq:model2}
	\begin{gathered}
		\frac{\partial\mathbf{U}}{\partial t} +  \nabla \cdot \mathcal{F}^{c}   + \mathbf{h} ~\nabla \cdot \mathbf{u} - \nabla \cdot \mathcal{F}^{v} - \mathbf{N}^{v} -  \mathbf{S} = \mathbf{0} \\ 
		\mathcal{F}^c  = (\mathbf{F}^c, \mathbf{G}^c), \quad \mathcal{F}^v  = (\mathbf{F}^v, \mathbf{G}^v) \\
		\mathbf{U} = \left[\begin{array}{c} \alpha_1 \\ \alpha_1 \rho_1 \\ \alpha_2 \rho_2 \\  \rho u \\ \rho v \\ \rho E \\\alpha_1 \rho_1 e_1 \\ \alpha_2 \rho_2 e_2 \end{array}\right], \quad
		\mathbf{F}^c = \left[\begin{array}{c} \alpha_1 u \\ \alpha_1 \rho_1 u \\  \alpha_2 \rho_2 u  \\  \rho {u^2} +  p  \\ \rho u v \\ (\rho E + p) u \\ \alpha_1 \rho_1 e_1 u \\ \alpha_2 \rho_2 e_2 u \end{array}\right], \quad
		\mathbf{G}^c = \left[\begin{array}{c} \alpha_1 v \\ \alpha_1 \rho_1 v \\  \alpha_2 \rho_2 v  \\  \rho u v  \\  \rho v^2 + p \\ (\rho E + p) v \\ \alpha_1 \rho_1 e_1 v \\ \alpha_2 \rho_2 e_2 v \end{array}\right], \quad
		\mathbf{h} = \left[\begin{array}{c} - \alpha_1 \\ 0 \\ 0 \\ 0 \\ 0 \\ 0 \\  \alpha_1 p_1  \\ \alpha_2 p_2 \\ \end{array}\right] \\
		\mathbf{F}^v = \left[\begin{array}{c} 0 \\ 0 \\ 0  \\  \tau_{xx}  \\ \tau_{yx} \\  \tau_{xx} u + \tau_{xy} v \\ \alpha_1 \left( \tau_{xx} u + \tau_{xy} v\right) \\  \alpha_2 \left( \tau_{xx} u + \tau_{xy} v\right) \end{array}\right], \quad
		\mathbf{G}^v = \left[\begin{array}{c}  0 \\ 0 \\ 0  \\  \tau_{xy}  \\ \tau_{yy} \\  \tau_{yx} u + \tau_{yy} v \\ \alpha_1 \left( \tau_{yx} u + \tau_{yy} v \right) \\  \alpha_2 \left( \tau_{yx} u + \tau_{yy} v \right)  \end{array}\right] \\
		\mathbf{N}^v = \left[\begin{array}{c} 0 \\ 0 \\ 0 \\ 0 \\ 0 \\ 0 \\ - \mathbf{u} \cdot \left( \nabla \cdot \left(  \alpha_1 \mathcal{\tau}\right) \right)  \\ - \mathbf{u} \cdot \left( \nabla \cdot \left(  \alpha_2 \mathcal{\tau}\right) \right) \\ \end{array}\right], \quad
		\mathbf{S} = \left[ \begin{array}{c}
			0 \\ 0 \\ 0 \\ \sigma \kappa \frac{\partial \alpha_1}{\partial x} + \rho g_x \\ \sigma \kappa \frac{\partial \alpha_1}{\partial y} + \rho g_y \\ \sigma \kappa \mathbf{u} \cdot \nabla \alpha_1 + \rho \mathbf{g} \cdot \mathbf{u}  \\ 0 \\ 0      
		\end{array} \right]  
	\end{gathered}
\end{equation}

 The Finite volume discretization of system without any relaxation terms can be written as
\begin{equation}\label{eq:model2_disc}
 \frac{\partial\mathbf{U}_i}{\partial t} + \frac{1}{\Omega_i} \left[  \sum^{4}_{l = 1} \mathcal{F}^{c}_{il} \cdot \mathbf{n}_{il} ~\Delta s_{il}  +  \mathbf{h}_i \sum^{4}_{l = 1}  {\left( u_{n} \right)}_{il} ~\Delta s_{il}  -  \sum^{4}_{l = 1} \mathcal{F}^{v}_{il} \cdot \mathbf{n}_{il} ~\Delta s_{il}\right]  -  \mathbf{N}^v_i  - \mathbf{S}_i   = \mathbf{0}
\end{equation}
Here, $\Omega_i$ denotes area of cell $i$ and ${\Delta s}_{il}$ represents the length and $n_{il}$ represents the normal face vector of the interface between cell $i$ and neighbouring cell $l$. In the subsequent sections the discretization of convective, viscous and surface tension terms are discussed in a detailed manner.

\subsubsection{Discretization of convective terms}

First we look into computation of conservative fluxes $\mathcal{F}^{c}$. With the rotation invariance property the normal conservative flux can be written as
\begin{equation}\label{eq:rotational1}
	\mathcal{F}^{c}_{il}  \cdot  \mathbf{n}_{il} = \mathrm{T}^{-1}_{il} \mathbf{F}^{c} \left( \mathbf{\hat{U}}_{L}, \mathbf{\hat{U}}_{R} \right). 
\end{equation}
where $\mathrm{T}^{-1}_{il}$ is the inverse rotational matrix and $  \mathbf{F}^{c} \left( \mathbf{\hat{U}}_{L}, \mathbf{\hat{U}}_{R} \right)$ is the normal conservative flux vector in locally rotated coordinate.  $\mathbf{\hat{U}}_{L}$  and $\mathbf{\hat{U}}_{R}$ are the set of normal conservative variables at the left and right side of the cell interface. The normal flux $  \mathbf{F}^{c} \left( \mathbf{\hat{U}}_{L}, \mathbf{\hat{U}}_{R} \right)$ at the cell interface is computed using Riemann solver. The expression for normal variables can be written as
\begin{equation}\label{eq:rotational2}
	\begin{gathered}
		\mathbf{\hat{U}}_{L/R} = \mathsf{T}_{il} \mathbf{U}_{L/R} = \left[ \begin{array}{c} \alpha_1 \\ \alpha_1 \rho_1 \\ \alpha_2 \rho_2 \\ \rho u_n \\ \rho u_t \\ \rho E \\ \alpha_1 \rho_1 e_1 \\ \alpha_2 \rho_2 e_2 \end{array} \right]_{L/R}, \quad 
		\mathrm{T}_{il} = \left[\begin{array}{cccccccc}
			1 & 0 & 0 & 0 & 0 & 0 & 0 & 0\\
			0 & 1 & 0 & 0 & 0 & 0 & 0 & 0\\
			0 & 0 & 1 & 0 & 0 & 0 & 0 & 0\\
			0 & 0 & 0 & n_{x} & n_{y} & 0& 0 & 0\\
			0 & 0 & 0 &-n_{y} & n_{x} & 0& 0 & 0\\
			0 & 0 & 0 & 0 & 0 & 1 & 0 & 0  \\
			0 & 0 & 0 & 0 & 0 & 0 & 1 & 0  \\
			0 & 0 & 0 & 0 & 0 & 0 & 0 & 1 
		\end{array}\right].  
	\end{gathered}
\end{equation}
Here, ${u}_n$ and ${u}_t$ are normal and tangential velocities at cell interface, and $(n_{x}, n_{y})$ are components of unit normal face vector $\mathbf{n}_{il}$. 

Besides the conservative fluxes, the non-conservative terms also contribute to the convective portion of the system, represented by $\mathbf{h} \nabla \cdot \mathbf{u}$ in Eq.~\eqref{eq:model2}. Its discretized counterpart, shown in Eq.~\eqref{eq:model2_disc}, requires the evaluation of $\mathbf{h}_i$ using cell-averaged variables, while the normal velocity at the cell faces $(u_n)_{il}$ is obtained using a Riemann solver.

\subsubsection{HLLC Riemann solver with capillary terms}\label{sec:Rimann_solver}
The normal flux vector, $\mathbf{F}^{c} \left( \mathbf{\hat{U}}_{L}, \mathbf{\hat{U}}_{R} \right)$, is determined using an HLLC solver. Although Riemann solvers incorporating capillary terms have already been developed in the literature \cite{fiveeq_st_css_2005,fiveeq_st_csf_2017,fiveeq_st_csf_2021}, they are primarily formulated for single-pressure models. For the two-pressure six-equation model, both an analysis of Generalised Riemann Invariants (GRI) and a Riemann solver that accounts for capillary effects are currently missing from the literature. The present study addresses this limitation. The eigenvalues and eigenvectors of the multiphase system with capillary terms, along with the Generalised Riemann Invariants analysis, are presented in \ref{sec:GRIs}.  

For the six-equation compressible multiphase flow model, the HLLC solver acts as an approximate Riemann solver characterized by two intermediate states separated by middle wave fields moving at a single speed $S^*$, bounded by outer waves with speeds $S_L$ and $S_R$. The normal flux vector, $\mathbf{F} \left( \mathbf{\hat{U}}_{L}, \mathbf{\hat{U}}_{R} \right)$, depends on the wave speeds relative to the interface and is determined according to the following conditions:
\begin{equation}
\mathbf{F} \left( \mathbf{\hat{U}}_{L}, \mathbf{\hat{U}}_{R} \right) = \mathbf{F} \left( \mathbf{U}^{int} \right), \quad \text{where} \quad \mathbf{U}^{int} =  
\begin{cases} 
	\mathbf{U}_{L}, & \text{if } 0 \leq S_L \\
	\mathbf{U}^{*}_{L}, & \text{if } S_L \leq 0 \leq S^{*} \\
	\mathbf{U}^{*}_{R}, & \text{if } S^{*} \leq 0 \leq S_R \\
	\mathbf{U}_{R}, & \text{if } S_R \leq 0
\end{cases}
\end{equation}

Using the Rankine-Hugoniot jump conditions, $\mathbf{F}(\mathbf{U}^{*}_K) - \mathbf{F}(\mathbf{U}_K) = S_K \left(\mathbf{U}^{*}_K - \mathbf{U}_K \right)$, across the left ($K = L$) and right ($K = R$) waves, the intermediate state variables, $\mathbf{U}^{*}_K$, are determined. Applying these relations to the mass, momentum, and total energy equations, and invoking the condition $u^{*}_{nR} = u^{*}_{nL} = S^{*}$ from Eq.~\eqref{eq:middle_wave}, we obtain:
\begin{equation}\label{eq:HLLCjump}
	\begin{aligned}
		\alpha^{*}_{jK}	\rho^{*}_{jK} & = 	\alpha_{jK} \rho_{jK} \left( \frac{S_K - u_{nK}}{S_K - S^{*}} \right)  \\
		p^{*}_K &= p_K + \rho_K (u_{nK}  - S_K ) ( u_{nK} - S^{*}) \\
		{E}^{*}_K &= \left(E_K+\left(S^{*}-u_{n K}\right)\left(S^{*}+\frac{p_K}{\rho_K\left(S_K-u_{n K}\right)}\right)\right)  \\
	\end{aligned}
\end{equation}.

Based on the Generalised Riemann Invariants (GRI) analysis performed in \ref{sec:GRIs}, the Laplace relation for the mixture pressure (Eq.~\eqref{eq:middle_wave}) across the contact wave is given by $p^{*}_R - p^{*}_L = \sigma \kappa \left(\alpha^{*}_{1R} - \alpha^{*}_{1L}\right)$. Furthermore, from Eqs.~\eqref{eq:left_wave} and \eqref{eq:right_wave}, we establish that $u^{*}_{tK} = u_{tK}$ and $\alpha^{*}_{K} = \alpha_{K}$. Applying these relations for the mixture pressure and volume fraction yields the contact wave speed:
\begin{equation}
	S^{*} = \frac{(p_R - p_L) + \rho_R u_{nR} (u_{nR} - S_R) - \rho_L u_{nL} (u_{nL} - S_L) - \sigma \kappa (\alpha_{1R} - \alpha_{1L} )}{\rho_R (u_{nR} - S_R) - \rho_L (u_{nL} - S_L)}.
\end{equation}
Evaluating this expression requires the curvature, $\kappa$, at the cell faces. This value can be approximated by a simple arithmetic average of the curvatures from the adjacent cell centers, i.e., $\kappa = \frac{1}{2}(\kappa_i + \kappa_l)$. Finally, the bounding wave speeds $S_L$ and $S_R$ are estimated using the following expressions \cite{davis1988simplified}:
\begin{equation}
	S_L = \min(u_{nL} - a_L, u_{nR} - a_R), \quad S_R = \max(u_{nL} + a_L, u_{nR} + a_R).
\end{equation}

For the phasic energy equations, the intermediate specific internal energy $e^{*}_{jK}$ is needed, which can be computed using the equation of state \eqref{eq:EOS}. This requires the phasic pressure $p^{*}_{jK}$ and density $\rho^{*}_{jK}$, which can be evaluated as~\cite{Saurel2009SimpleAE}:
\begin{equation}\label{eq:Prel1}
	p^{*}_{jK} = (p_{jK} + \pi_j) \frac{(\gamma_j -1)\rho_{jK} - (\gamma_j + 1)\rho^{*}_{jK} }{(\gamma_j -1)\rho^{*}_{jK} - (\gamma_j + 1)\rho_{jK}} - \pi_j.  
\end{equation}
Here, $\rho^{*}_{jK}$ is taken from Eq.~\eqref{eq:HLLCjump}. We can also consider the expression for $p^{*}_{jK}$ given by the isentropic relation, which can be written as:
\begin{equation}\label{eq:Prel2}
	p^{*}_{jK} = \left(p_{jK} + \pi_j\right) \left(\frac{\rho^{*}_{jK}}{\rho_{jK}}  \right)^{\gamma} - \pi_j .  
\end{equation}
For moderate density ratios, both formulations yield the same results; differences are only observed in the presence of high density ratios. Specifically, the expression in Eq.~\eqref{eq:Prel1} can produce large negative values. Thus, we can use the relation in Eq.~\eqref{eq:Prel2} to avoid any computational breakdown in cases involving high-strength shock waves. For a detailed discussion, please refer to \ref{sec:failure}.

\subsubsection{Surface tension terms}
The surface tension terms appearing in the momentum equation require the computation of the volume fraction gradient, $\nabla \alpha_1$. Using the Green–Gauss method, these terms are evaluated as
\begin{equation}\label{eq:st_mom}
	\begin{split}
		&\sigma \kappa \frac{\partial \alpha_1}{\partial x} = \sigma \kappa_i \frac{1}{\Omega_i} \sum^{4}_{l = 1} {(\alpha_1)}_{il} ~{(n_x)}_{il} ~\Delta s_{il}, \\ 
		&\sigma \kappa \frac{\partial \alpha_1}{\partial y} = \sigma \kappa_i ~\frac{1}{\Omega_i} \sum^{4}_{l = 1} {(\alpha_1)}_{il} ~{(n_y)}_{il} ~\Delta s_{il}
	\end{split}
\end{equation}
Surface tension term appearing in the total energy can be re-expressed as \cite{fiveeq_st_csf_2017,fiveeq_st_csf_2021} 
\begin{equation}
	\sigma \kappa \mathbf{u} \cdot \nabla \alpha_1 = \sigma \kappa \left( \nabla \cdot (\alpha_1 \mathbf{u})  - \alpha_1 \nabla \cdot (\mathbf{u})  \right)
\end{equation}
Since the above expression is in divergence form it can be discretized as
\begin{equation}\label{eq:st_ene}
	\sigma \kappa \left( \nabla \cdot (\alpha_1 \mathbf{u})  - \alpha_1 \nabla \cdot (\mathbf{u})  \right) = \frac{1}{\Omega_i} \sigma \kappa_i \left( \sum^{4}_{l = 1} (\alpha_1)_{il} (u_n)_{il} ~\Delta s_{il} - (\alpha_1)_{i} \sum^{4}_{l = 1} (u_n)_{il} ~\Delta s_{il} \right)
\end{equation}
To maintain consistency with the convective fluxes \cite{fiveeq_st_csf_2017, garrick2017interface,fiveeq_st_csf_2021}, the interface quantities ${\left(\alpha_1 \right)}_{il} $ and ${\left(u_n\right)}_{il}$ in Eqs.~\eqref{eq:st_mom} and \eqref{eq:st_ene} are obtained from the solution of the Riemann problem described in Section \ref{sec:Rimann_solver}. 

\subsubsection{Curvature computation}
The surface tension formulation requires the computation of the interface curvature $\kappa$ at the cell center, given by
\begin{equation}\label{eq:curvature}
	\kappa = - \nabla \cdot \mathbf{n}^{\mathrm{I}}, \quad \mathbf{n}^{\mathrm{I}} = - \frac{\nabla {\psi}}{|\nabla {\psi}|} .
\end{equation}
Here, $\mathbf{n}^{\mathrm{I}}$ is interface normal vector, which is computed using the smoothed field, $\psi$. The smoothed field is obtained from the volume fraction field through the convolution process defined as
\begin{equation}
	{\psi}_i = \sum_{m} {(\alpha_1)}_m ~K(|\mathbf{r}_l - \mathbf{r}_i|, \delta)~ \Omega_{m}.
\end{equation}
Following \cite{bhat2019contact,sun2010coupled}, smoothing kernel given by cubic spline function \cite{monaghan1992smoothed} is used, which is defined as  
\begin{equation}
	K(r,\delta) =
	\begin{cases}
		\dfrac{40}{7\pi}
		\left(
		1 - 6\left(\dfrac{r}{\delta}\right)^2
		+ 6\left(\dfrac{r}{\delta}\right)^3
		\right),
		& \text{if } \dfrac{r}{\delta} < \dfrac{1}{2}, \\[10pt]
		
		\dfrac{80}{7\pi}
		\left(
		1 - \dfrac{r}{\delta}
		\right)^3,
		& \text{if } \dfrac{1}{2} \le \dfrac{r}{\delta} < 1, \\[10pt]
		
		0,
		& \text{otherwise}.
	\end{cases}
\end{equation}
The parameter $\delta$ in above expression defines width of the region used for smoothing and it is generally selected as $\delta = 3 H$ \cite{bhat2019contact,sun2010coupled}, where $H$ is a measure of grid size. The computation of curvature from the smoothed volume fraction field requires the estimation of gradients. Due to its robustness and applicability to arbitrary meshes, the weighted least squares method is employed to evaluate the gradients required for curvature calculation.  

The truncated Taylor series expansion at the neighboring cell center $l$ about cell center $i$ can be written as
\begin{equation}\label{eq:taylor}
	f_l = f_i + \left.\dfrac{\partial f}{\partial x}\right|_{i} ( x_{l} - x_i) + \left.\dfrac{\partial f}{\partial y}\right|_{i} ( y_{l} - y_i) + \mathcal{O}(\Delta x^2, \Delta y^2)
\end{equation} 
Using above Eq.~\eqref{eq:taylor} for a set of neighbors, we get following linear system. 
\begin{equation}\label{eq:wls1}
	\underbrace{\begin{bmatrix}
	  w_1 \left( x_{1} - x_i \right) & w_1 \left( y_{1} - y_i \right) \\
	\vdots     & \vdots     \\
 w_l \left(	x_{l} - x_i \right) &  w_l \left( y_{l} - y_i \right)  \\
	\vdots     & \vdots    \\
 w_N \left(	x_{N} - x_i \right)  &   w_N \left( y_{N} - y_i \right)
	\end{bmatrix}}_{\mathrm{S}}
\underbrace{\begin{bmatrix}
		\left.\dfrac{\partial f}{\partial x}\right|_{i} \\[12pt]
		\left.\dfrac{\partial f}{\partial y}\right|_{i} 
	\end{bmatrix}}_{ \mathbf{df} }
	=
\underbrace{\begin{bmatrix}
		  w_1 \left( f_{1} - f_i \right) \\
		\vdots \\
	  w_l \left(	f_{l} - f_i \right) \\
		\vdots \\
	  w_N \left(	f_{N} - f_i \right) 
	\end{bmatrix}}_{ \mathbf{ \Delta f} }
\end{equation}
Here, $w_l$ are distance based weights defined as $w_l = \frac{1}{{\left( x_{l} - x_i \right)}^2 + { \left( y_{l} - y_i \right)}^2}$. The linear system in Eq.~\eqref{eq:wls1} is constructed for all neighboring cells surrounding cell $i$, including indirect neighbors, as illustrated in \fig{fig:FVM1} (a). The resulting system is overdetermined and is solved using the least-squares method. In the present work, the classical normal equation approach is used, which can be written as
\begin{equation}\label{eq:wls2}
 \mathbf{df} = {\left(\mathrm{S}^T \mathrm{S}\right)}^{-1} \mathrm{S}^T \mathbf{ \Delta f}.
\end{equation}

To calculate the curvature, the derivatives of the smoothed function, $(\frac{\partial \psi}{\partial x}, \frac{\partial \psi}{\partial y})$, are first evaluated using the weighted least-squares method detailed above. The interface normal vector is then obtained through normalization as $\mathbf{n}^{\mathrm{I}} = \frac{\nabla {\psi}}{|\nabla {\psi}|} $. Using this computed normal, the derivatives $(\frac{\partial n^\mathrm{I}_x}{\partial x}, \frac{\partial n^\mathrm{I}_y}{\partial y})$ are subsequently evaluated by solving the same least squares system with $f=n^\mathrm{I}_x$ and $f=n^\mathrm{I}_y$, respectively. Finally, the curvature is determined as $\kappa = - \left( \frac{\partial n^\mathrm{I}_x}{\partial x} + \frac{\partial n^\mathrm{I}_y}{\partial y} \right) $.

The computed curvature varies along the interface normal, with numerical errors increasing significantly near the pure-phase regions. To reduce the error in curvature estimation, an iterative correction procedure \cite{fiveeq_st_csf_2017,garrick2017interface,furfaro2020towards} is used. The corrected curvature after the $(n+1)$-th iteration is given by
\begin{equation}\label{eq:curv_corr}
	\kappa^{n+1}_i = \frac{w_i \kappa^{n}_i + \sum_{l} w_l \kappa^{n}_l  }{w_i + \sum_{l} w_l }.
\end{equation}
The stencil of neighboring cells used in Eq.~\eqref{eq:curv_corr} is identical to that employed for curvature estimation (shown in \fig{fig:FVM1}~(a)). Here, the weights are defined as $w = \left(\psi (1 - \psi)  \right) $, which attains its maximum value at $\psi = 0.5$. Consequently, the iterative procedure redistributes the curvature evaluated near $\psi = 0.5$ across the neighboring cells, thereby improving the overall curvature field. The number of iterations used in the method is problem specific and it is reported in the test cases.

\subsubsection{Viscous terms}
 
The viscous fluxes appearing in the momentum and mixture total energy equations of the multiphase model \eqref{eq:model} are similar to those in the Navier–Stokes equations. The only additional contribution arises in the phasic energy equations, where the viscous dissipation term $\alpha_j \mathsf{\tau} : \nabla \mathbf{u}$ appears. This term is re-expressed as \cite{six_eq_viscous} $\nabla (\alpha_j \mathsf{\tau} \cdot \mathbf{u}) - \mathbf{u} \cdot \left(\nabla \cdot (\alpha_j \mathsf{\tau}) \right) $, and incorporated into the formulation given in Eq.~\eqref{eq:model2}. 

For convenience, the viscous terms are grouped into the tensor $\mathcal{F}^{v}$ and the vector $\mathbf{N}^{v}$. The divergence term $\nabla \cdot \mathcal{F}^{v}$ in the system \eqref{eq:model2} can be directly discretized as
\begin{equation}\label{eq:viscous1}
	\sum^{4}_{l = 1} \mathcal{F}^{v}_{il} \cdot \mathbf{n}_{il} ~\Delta s_{il} = 	\sum^{4}_{l = 1} \left( \mathbf{F}^{v}_{il} {(n_x)}_{il} + \mathbf{G}^{v}_{il} {(n_y)}_{il} \right) ~\Delta s_{il}.
\end{equation}
The only expressions appearing in the vector $\mathbf{N}^{v}$ are the terms $- \mathbf{u} \cdot \left(\nabla \cdot (\alpha_j \mathsf{\tau}) \right)$ can be discretized as
\begin{equation}\label{eq:viscous2}
	\begin{gathered}
		\mathbf{u} \cdot \left(\nabla \cdot (\alpha_j \mathsf{\tau}) \right) = \frac{1}{\Omega_i} \left[ {u}_i 	\sum^{4}_{l = 1} {(\alpha_j)}_{il} \left( (\tau_{xx})_{il} {(n_x)}_{il} + (\tau_{xy})_{il} {(n_y)}_{il} \right) ~\Delta s_{il} \right] \\
		+ \frac{1}{\Omega_i} \left[{v}_i	\sum^{4}_{l = 1} {(\alpha_j)}_{il} \left( (\tau_{yx})_{il} {(n_x)}_{il} + (\tau_{yy})_{il} {(n_y)}_{il} \right) ~\Delta s_{il} \right].
	\end{gathered}
\end{equation}
The evaluation of the viscous stress tensor $\mathsf{\tau}$ at the cell faces requires velocity gradients $\nabla u, \nabla v$. These gradients are computed at the cell faces using the Green–Gauss method applied to a diamond-shaped control volume (shown in \fig{fig:FVM1}~(b)), commonly referred to as the Coirier diamond \cite{coirier1994adaptively}. The gradient of a scalar variable $\phi$ at the cell face shared by cell $i$ and $l$ can be written as
\begin{equation}
	\nabla \phi = \frac{1}{2\Omega_{iplq}} \left[\left({\phi_i + \phi_p }\right)\mathbf{n}_{ip}~\Delta s_{ip} + \left({\phi_p + \phi_l }\right) \mathbf{n}_{pl}~\Delta s_{pl} + \left({\phi_l + \phi_q }\right)\mathbf{n}_{lq}~\Delta s_{lq} + \left({\phi_q + \phi_i }\right)\mathbf{n}_{qi}~\Delta s_{qi} \right].
\end{equation}
In the above expression, the values at vertices $p$ and $q$ are also required. These are obtained as area-weighted averages of the cell-centered values of the cells sharing the respective vertex. For example scalar value at vertex $p$ can be defined 
\begin{equation}
	\phi_p = \frac{\sum^{N}_{r=1} \Omega_r \phi_r }{\sum^{N}_{r=1} \Omega_r}.
\end{equation}
Where $N$ is the total number of cells sharing a common vertex $p$. 

In addition to the gradients, the evaluation of the diffusive fluxes requires the viscosity $\mu$, volume fraction $\alpha_j$, and velocity $\mathbf{u}$ at the cell faces. These quantities are computed by averaging the values from the neighboring cells sharing the face.  
 
\subsection{Relaxation step}\label{sec:relax_step}
The solution obtained from the evolution step is in a non-equilibrium state. It is subsequently projected onto mechanical equilibrium during the relaxation step by using instantaneous pressure relaxation (in the limit $K_p \rightarrow \infty$ ) \cite{Saurel2009SimpleAE, zein2010, liquidgas2013, pelanti2014mixture}. Retaining only the relaxation terms in model \eqref{eq:model} leads to following system
\begin{equation}\label{eq:relax_ODE}
	\begin{gathered}
		\frac{d}{dt}\left[\begin{array}{c} \alpha_1 \\ \alpha_1 \rho_1 \\ \alpha_2 \rho_2 \\  \rho u \\ \rho v \\ \rho E \\\alpha_1 \rho_1 e_1 \\ \alpha_2 \rho_2 e_2 \end{array}\right] = \left[\begin{array}{c}  K_p \left(p_1 - p_2  \right)  \\ 0 \\ 0 \\0\\ 0 \\ 0\\ -K_p p_I \left(p_1 - p_2 \right) \\ K_p p_I \left(p_1 - p_2 \right)\end{array}\right]  .
	\end{gathered}
\end{equation}
Based on the system of ODEs (Eq.~\eqref{eq:relax_ODE}), the product of the volume fraction and the density ($\alpha_j \rho_j$) and the velocity components ($u$ and $v$) remain constant during the relaxation step. By substituting the volume fraction equation into the phasic energy equations, the system in Eq.~\eqref{eq:relax_ODE} reduces to
\begin{equation}\label{eq:relax_ODE2}
	\begin{gathered}
		\frac{d {\left( \alpha_j \rho_j e_j \right)}}{dt} = -p_I \frac{d {\alpha_j }}{dt}, \quad j = 1, 2.
	\end{gathered}
\end{equation}
Above ODEs can be discretized as
\begin{equation}\label{eq:relax_eq2}
	\begin{gathered}
		{\left( \alpha_j \rho_j e_j \right)}^{*} - {\left( \alpha_j \rho_j e_j \right)}^{o} = - \bar{p}_I {\left( \alpha^{*}_j  - \alpha^{o}_j \right) } , \quad j = 1,2.
	\end{gathered}
\end{equation}
The superscripts $o$ and $*$ denote the variables before and after the relaxation step, respectively. Here, $\bar{p}_I$ represents the numerical approximation of $\frac{1}{\left( \alpha^{*}_j  - \alpha^{o}_j \right)} \int p_I d{\alpha_j }$, and it can be taken as $p^{o}_I$, $p^{*}_I$, or $\frac{p^{*}_I + p^{o}_I}{2}$ \cite{saurel2001, Saurel2009SimpleAE,pelanti2014mixture}. It is observed that in the absence of a strong shock, any of these choices yields similar results. Assuming $\bar{p}_I = p^{o}_I$, and with the help of the SGEOS (Eq. \eqref{eq:EOS}), Eq. \eqref{eq:relax_eq2} becomes:
\begin{equation}\label{eq:relax_eq3}
	\begin{gathered}
\alpha^{*}_j  \left( p^{*}_j + \gamma_j \pi_j \right) - \alpha^{o}_j  \left( p^{o}_j + \gamma_j \pi_j \right) = - p^{o}_I \left( \alpha^{*}_j - \alpha^{o}_j\right) \left( \gamma_j -1\right).
	\end{gathered}
\end{equation}

Since, the pressure relaxation is done by considering instantaneous relaxation, the pressures $p^{*}_1, p^{*}_2$ after relaxation follows mechanical equilibrium condition. In the case of surface tension it would be $p^{*}_1 - p^{*}_2 = \sigma \kappa$ \cite{saurel1999, allmach_pressure_2021, panchal2023seven}. Substituting the conditions in Eq.\eqref{eq:relax_eq3} with volume fraction constraint $\sum_j \alpha_j = 1$ we get single quadratic equation as
\begin{equation}\label{eq:relax_eq4}
	\begin{split}
	&\boxed{A_1 {(p^{*})}^2 + A_2 {(p^{*})} + A_3 = 0,\ \ } \\
\text{Where,}& \\
&A_1 =  1,\\
&A_2 = \left( \alpha^{o}_1 C_2 + \alpha^{o}_2 C_1\right) - \alpha^{o}_1 \left( p^{o}_1 - \sigma \kappa \right)  - \alpha^{o}_2 \left( p^{o}_2  - \sigma \kappa \right), \\ 
&A_3 = -\left( \alpha^{o}_1 (p^{o}_1 - \sigma \kappa) C_2 + \alpha^{o}_2 p^{o}_2  (C_1 + \sigma \kappa)\right), \\
&C_j = \gamma_j \pi_j + p^{o}_I (\gamma_j -1).
	\end{split}
\end{equation}
Keeping the positive root as a solution for $p^{*}$:
\begin{equation}
	p^{*} = \frac{-A_2 + \sqrt{A_2^2 - 4 A_1 A_3}}{2 A_1},
\end{equation}
values of the phasic pressure are set as $p^{*}_1 = p^{*} + \sigma \kappa$ and $p^{*}_2 = p^{*}$. Volume fraction $\alpha^{*}_j$ after relaxation are computed from Eq.~\eqref{eq:relax_eq3} and phasic densities are computed as $\rho^{*}_j = \frac{\left(\alpha_j \rho_j\right)^o}{\alpha^{*}_j}$.
 
In order to enforce the conservation of mixture total energy, the pressure $p_j$ and energy $e_j$ values obtained from pressure relaxation are corrected using mixture total energy equation \cite{Saurel2009SimpleAE,zein2010,overbee}. The corrected pressure is obtained using mixture equation of state. Using the surface tension constraint $p^{**}_1 - p^{**}_2 = \sigma \kappa$ and SGEOS~\eqref{eq:EOS} relation final pressures are obtained as 
\begin{equation}\label{eq:mixtureEOS}
\quad p^{**}_1 = p^{**}_2 + \sigma \kappa, \quad p^{**}_2 = \frac{(\rho e)^{o} - \left(\sum_j \frac{\alpha^{*}_j \gamma_j \pi_j}{\gamma_j - 1} + \frac{\alpha^{*}_1 \sigma \kappa}{\gamma_1 - 1}  \right)}{\sum_j \frac{\alpha^{*}_j}{\gamma_j - 1}}
\end{equation}

\subsection{Second order formulation}\label{sec:second_order}
To obtain a second-order spatial discretization, the set of primitive variables at the cell faces, $\mathbf{W}_{L/R} = [\alpha_1, \rho_1, \rho_2, u, v, p_1, p_2]_{L/R}$ is reconstructed from the neighboring cell-averaged quantities $(\mathbf{W}_i, \mathbf{W}_l)$. The reconstruction procedure is based on a Taylor series expansion truncated after the first-order terms, and can be written as
\begin{equation}\label{eq:reconstruct}
	\begin{aligned}
		{W}_L &= {W}_i 
		+ \left[
		\left. \frac{\partial {W}}{\partial x} \right|_i (x_{il} - x_i)
		+ 
		\left. \frac{\partial {W}}{\partial y} \right|_i (y_{il} - y_i)
		\right], \\[6pt]
	   {W}_R &= {W}_l 
		+ \left[
		\left. \frac{\partial {W}}{\partial x} \right|_l (x_{il} - x_l)
		+ 
	    \left. \frac{\partial {W}}{\partial y} \right|_l (y_{il} - y_l) \right].
	\end{aligned}
\end{equation}
In this work, the gradients $ \frac{\partial W}{\partial x}, \frac{\partial W}{\partial y} $ for all primitive variables, except the volume fraction, are computed using the SWDLS method \cite{SDWLS, mandal2014high,MANDAL2015669}. The method uses the weighted least-squares formulation described in Eqs.~\eqref{eq:wls1} and \eqref{eq:wls2}. The only modification is the use of solution-dependent weights defined as $w_l = \frac{1}{\left|W_l - W_i\right| + \epsilon}$, where $\epsilon$ is a small positive number added to avoid division by zero.

Only for the volume fraction $\alpha_1$, the interface-sharpening reconstruction proposed by \citet{overbee} is used. In this approach, the compressive Overbee limiter, is used together with the gradients computed by weighted least square method with the distance based weights. The use of this dedicated reconstruction procedure for the volume fraction effectively reduces numerical smearing of the interface~\cite{overbee,nguyen2022fully,furfaro2020towards}.

For second-order temporal accuracy, the strong stability-preserving Runge-Kutta (SSPRK) method is employed. Representing the evolution step (Section \ref{sec:evol_step}) and the relaxation step (Section \ref{sec:relax_step}) with the operators $\mathcal{L}^{\Delta t}_{E}$ and $\mathcal{L}^{\Delta t}_{R}$, respectively, the implementation of the second-order SSPRK scheme is written as
\begin{equation}
	\begin{aligned}
		\mathbf{U}^{(1)} &= \mathcal{L}^{\Delta t}_{R}\mathcal{L}^{\Delta t}_{E}\left(\mathbf{U}^n\right), \\[6pt]
		\mathbf{U}^{(2)} &= \mathcal{L}^{\Delta t}_{R}\mathcal{L}^{\Delta t}_{E}\left( \mathbf{U}^{(1)} \right), \\[6pt]
		\mathbf{U}^{(3)} &= \frac{1}{2} \left( \mathbf{U}^n + \mathbf{U}^{(2)} \right), \\[6pt]
		\mathbf{U}^{n+1} &= \mathcal{L}^{\Delta t}_{R} \left( \mathbf{U}^{(3)} \right).
	\end{aligned}
\end{equation}
 
 \begin{figure}[H]
 	\centering
 	\includegraphics[width=0.65\textwidth]{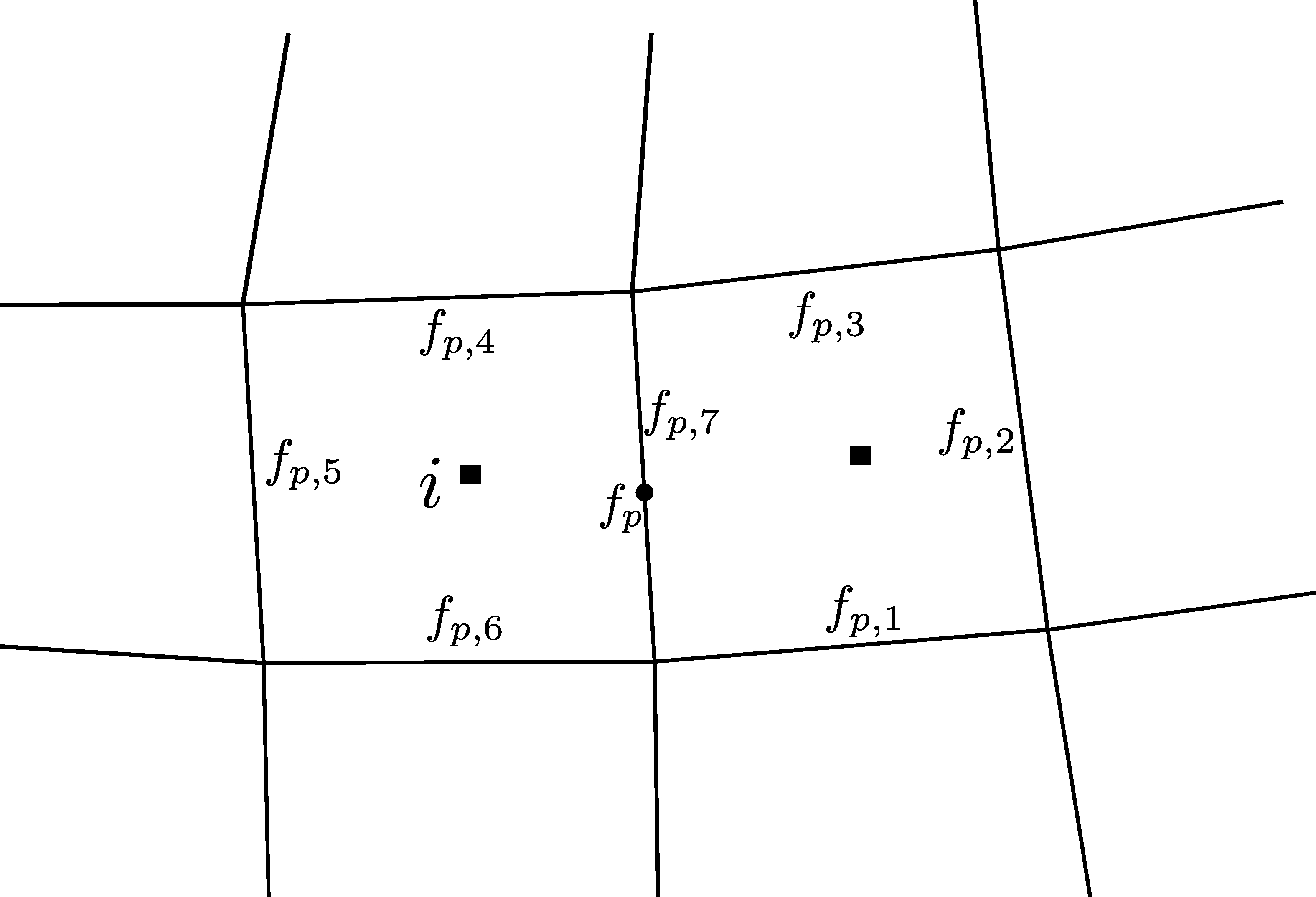}
 	\caption{Schematic illustrating the local pressure functions, $f_{p, r}$, required to evaluate the pressure function, $f_p$, at the face shared by cells $i$ and $l$.}
 	\label{fig:Psensor}
 \end{figure}
 
\section{Low Mach correction with modified scaling factor}\label{sec:LowMach}
Standard Riemann solvers are known to produce inaccurate results in the low-Mach regime. To mitigate this issue, a low-Mach correction approach originally developed for single-phase flows \cite{thornber2008improved, rieper2011low, osswald2016l2roe, gogoi2025enhanced, gogoi2023low} was extended to multiphase flows in our recent work \cite{bharate2025enhanced}. The effectiveness of this method has been established through various numerical experiments and asymptotic analyses~\cite{bharate2025enhanced}. In this approach, the velocity differences in the convective flux expressions are scaled by the local Mach number. To implement this correction, the following reconstructed left and right velocities \cite{thornber2008improved} are used in place of the standard left and right velocities, $(u_{nL}, u_{nR}, u_{tL}, u_{tR})$, in the HLLC flux expressions presented in the Section~\ref{sec:Rimann_solver}.
\begin{equation}
	\begin{gathered}
			{u}^{r}_{nL} = \frac{u_{nR} + u_{nL}}{2} - f_{\text{lm}} \frac{u_{nR} - u_{nL}}{2}, \quad {u}^{r}_{tL} = \frac{u_{tR} + u_{tL}}{2} - f_{\text{lm}} \frac{u_{tR} - u_{tL}}{2} \\
		{u}^{r}_{nR} = \frac{u_{nR} + u_{nL}}{2} + f_{\text{lm}} \frac{u_{nR} - u_{nL}}{2}, \quad {u}^{r}_{tR} = \frac{u_{tR} + u_{tL}}{2} + f_{\text{lm}} \frac{u_{tR} - u_{tL}}{2}
	\end{gathered}
\end{equation}
In our previous work \cite{bharate2025enhanced}, the scaling factor, $f_{\text{lm}}$, was defined directly as the local Mach number. At the cell face, this is computed as:
\begin{equation}\label{eq:flm}
	 f_{\text{lm}} = f_M, \quad f_M = \min\left(1, \max\left( \frac{\sqrt{u^2_{nL} + u^2_{tL}}}{a_L}, \frac{\sqrt{u^2_{nR} + u^2_{tR}}}{a_R} \right)    \right).
\end{equation}

While the low-Mach correction effectively reduces excessive diffusion in the low-Mach limit ($M \rightarrow 0$), it is prone to producing instabilities in the vicinity of pressure jumps. To circumvent this problem, the present study introduces a modified scaling factor, $f_{\text{lm}}$, which is evaluated as follows:
\begin{equation}\label{eq:modified_flm}
	f_{\text{lm}} = f_M (1 - fp) + fp. 
\end{equation}
Here, $f_p$ is a multidimensional pressure function that detects the presence of pressure jumps and reduces the effect of the low-Mach correction based on the strength of the pressure jump. The value of $f_p$ at the cell face shared by cells $i$ and $l$ is computed by evaluating the local pressure function, $f_{p,r}$, across all faces of the adjacent cells $i$ and $l$, as illustrated in \fig{fig:Psensor}. The pressure functions $f_p$ and $f_{p,r}$ are defined as follows \cite{gogoi2023low}:
\begin{equation}\label{eq:p_sensor}
	{f_p} = \min_r(f_{p,r}), \quad f_{p,r} = \sqrt{ \frac{| p_R - p_L |  }{\max(p_L, p_R)} }.
\end{equation}

The pressure function based approach has also been employed in the single-phase flow studies \cite{zhang2017robust, gogoi2023low, HLLC_IJNMF, gogoi2025enhanced} to mitigate shock instabilities in the low-Mach regime. However, in multiphase flows, pressure jumps can be induced by other factors such as surface tension at the phase interface or by hydrostatic pressure gradients within high-density fluids. Our numerical experiments demonstrate that our modified scaling factor, $f_{\text{lm}}$, enhances stability for low-Mach multiphase flows ($M \rightarrow 0$).

\section{Results}\label{sec:results}
To assess the capability of the proposed numerical method, various numerical experiments were conducted for a wide range of problems. These include the dam-break, static droplet, oscillating droplet, and rising bubble test cases. These problems are low-Mach test cases which include surface tension and viscous effects. Additionally, to test the robustness of the numerical method for high-speed flows, we simulate a shock-induced bubble collapse. Numerical results for all the test cases were obtained using a second-order scheme on a structured grid. A CFL number of 0.8 is used for the test cases unless otherwise specified.
 
\subsection{Dam-break}
The classic dam-break test case involves the collapse of a water column under gravity. Since the water starts moving from rest and the Mach number remains below  $10^{-2}$ throughout the simulation, the low-Mach correction is essential for maintaining numerical accuracy. The effectiveness of this correction has already been demonstrated in recent work \cite{bharate2025enhanced}. It is well known that viscous and capillary effects do not significantly alter the results of the dam-break problem. Therefore, the primary objective of this test case is to examine the effect of the modified scaling factor, $f_M (1 - f_p) + f_p$. The geometric configuration and mesh resolution are identical to those used in the previous study \cite{bharate2025enhanced}. The material properties of water and air employed in the present simulations are given by
 \begin{equation}
 	\begin{split}
 		&\rho_1 = 1000 ~\text{kg/m}^3,  \quad \mu_1 = 10^{-3} ~\text{Pa s}, \quad \gamma_1 = 4.4, \quad \text{and}~\pi_1 = 6 \times 10^8 ~\text{Pa}. \\
 		&\rho_2 = 1 ~\text{kg/m}^3, \quad \mu_2 = 1.8 \times 10^{-5} ~\text{Pa s},  \quad \gamma_2 = 1.4, \quad \text{and}~\pi_2 = 0 ~\text{Pa}, \\
 		& \sigma = 0.072 \text{N/m}, \quad \mathbf{g} = (0, -9.8) \text{m/s}^2.
 	\end{split}
 \end{equation}

\begin{figure}[H]
 	\centering
 	\includegraphics[width=\textwidth]{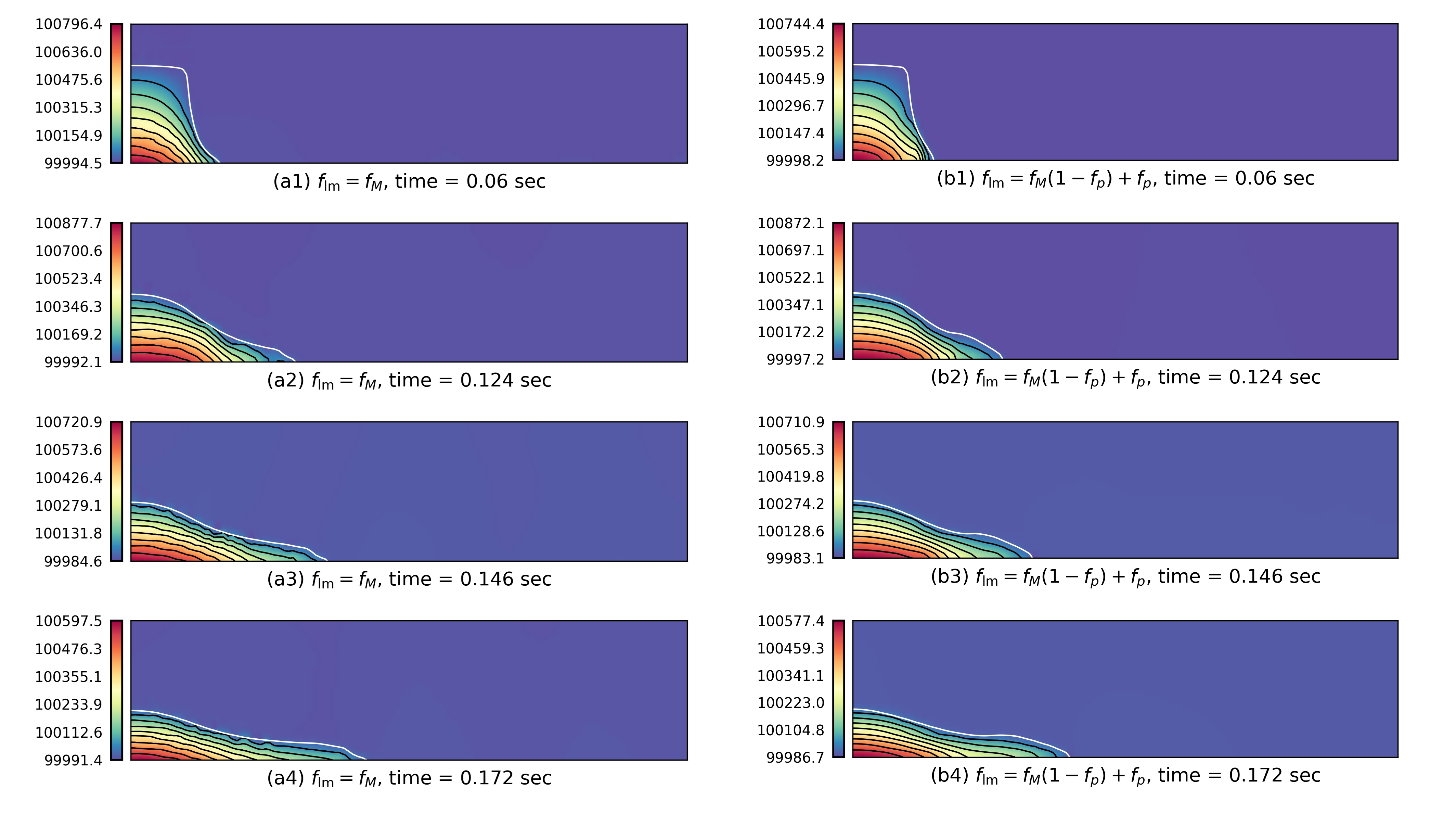}
 	\caption{Pressure contours and interface movement for the dam break problem. The black lines denote pressure contours, and the white line represents the interface ($\alpha = 0.5$). The left-hand plots show the results obtained using the scaling factor $f_M$, while the right-hand plots show the results obtained using the modified scaling factor $f_M (1 - f_p) + f_p$. }
 	\label{fig:dbcontours}
 \end{figure}

The pressure contours along with the interface profiles at various time instants for the numerical results obtained with the different scaling factors, $f_M$ and $f_M \left(1 - f_p\right) + f_p$, are shown in \fig{fig:dbcontours}. In the left-hand plots, the black contour lines reveal spurious pressure oscillations, particularly near the interface (white line). Since the pressure variation due to the hydrostatic effect in the water side is higher, the low-Mach correction with the simple scaling factor $f_M$ results in instability. However, the modified scaling factor $f_M \left(1 - f_p\right) + f_p$ rectifies the problem by reducing the effect of the low-Mach correction in regions of high pressure variation. The effectiveness of the modified factor can be observed in the right-side plots of \fig{fig:dbcontours}.

\begin{figure}[H]
 	\centering
 	\includegraphics[width=\textwidth]{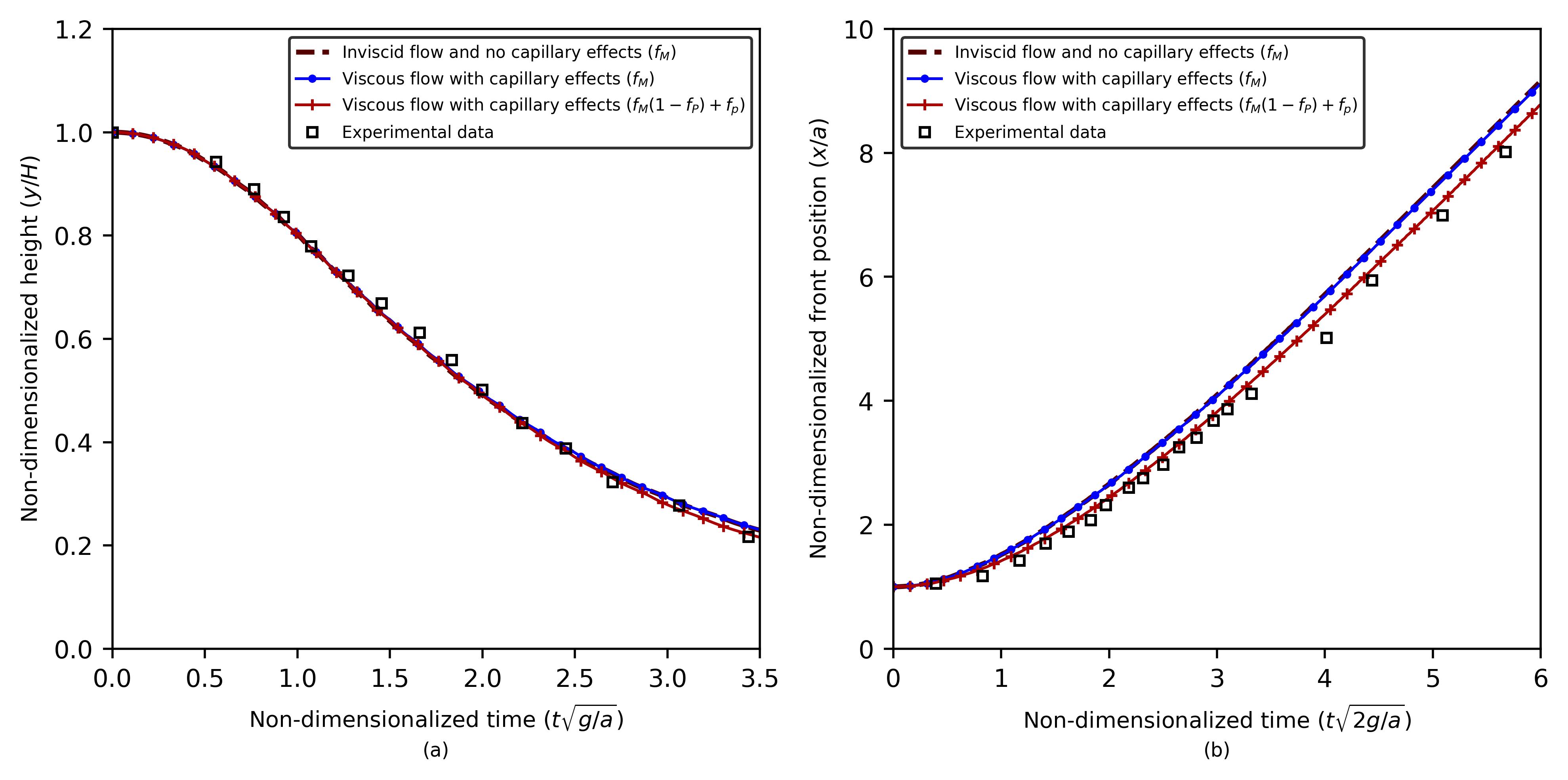}
 	\caption{Comparison of the numerical results, obtained using two different scaling factors, against the experimental data of \citet{martin1952experimental}: (a) non-dimensional water column height versus time, and (b) non-dimensional front position versus time.}
 	\label{fig:dbcompare}
 \end{figure}

For a quantitative comparison, time histories of the water column height and front position for the different numerical results, along with the experimental data \cite{martin1952experimental}, are shown in \fig{fig:dbcompare}. The comparison shows that the inclusion of viscous and capillary effects has a negligible influence on the results. The presence of spurious pressure oscillations also affects the interface movement and the overall accuracy of the numerical results, which can be observed in \fig{fig:dbcompare}. With the modified scaling factor, the numerical results agree more closely with the experimental data. For example, at the non-dimensional time $(t\sqrt{2g/a})$ = 4.43, the percentage error in the front position decreases from 8.22\% to 2.85\% after applying the modified scaling factor.

\subsection{Static droplet}\label{sec:Static} 
To validate the numerical scheme's efficacy in capturing the surface tension-induced pressure jump across the phase interface, the standard static droplet benchmark is simulated. In this problem, a circular liquid droplet of radius $R = 0.2$~m, centered at $(0.5, 0.5)$ inside a unit square domain, is considered. All boundaries of the computational domain are treated as walls. Since a uniform curvature is desired in this test, the curvature correction procedure (Eq.~\eqref{eq:curv_corr}) is applied using 10 iterations. The properties of the liquid (phase 1) and the surrounding air (phase 2) are defined as follows \cite{seveneq_st_path_2015}.
\begin{equation}
	\begin{split}
		&\rho_1=1000~\text{kg/m}^3, \quad \gamma_1=2.4, \quad \text{and}~\pi_1=10^7~\text{Pa}, \\
		&\rho_2=1~\text{kg/m}^3, \quad \gamma_2=1.4 \quad \text{and}~\pi_2=0~\text{Pa}, \\
		&\sigma = 200 \text{N/m}.
	\end{split}
\end{equation}

\begin{figure}[H]
	\centering
	\includegraphics[scale = 0.5]{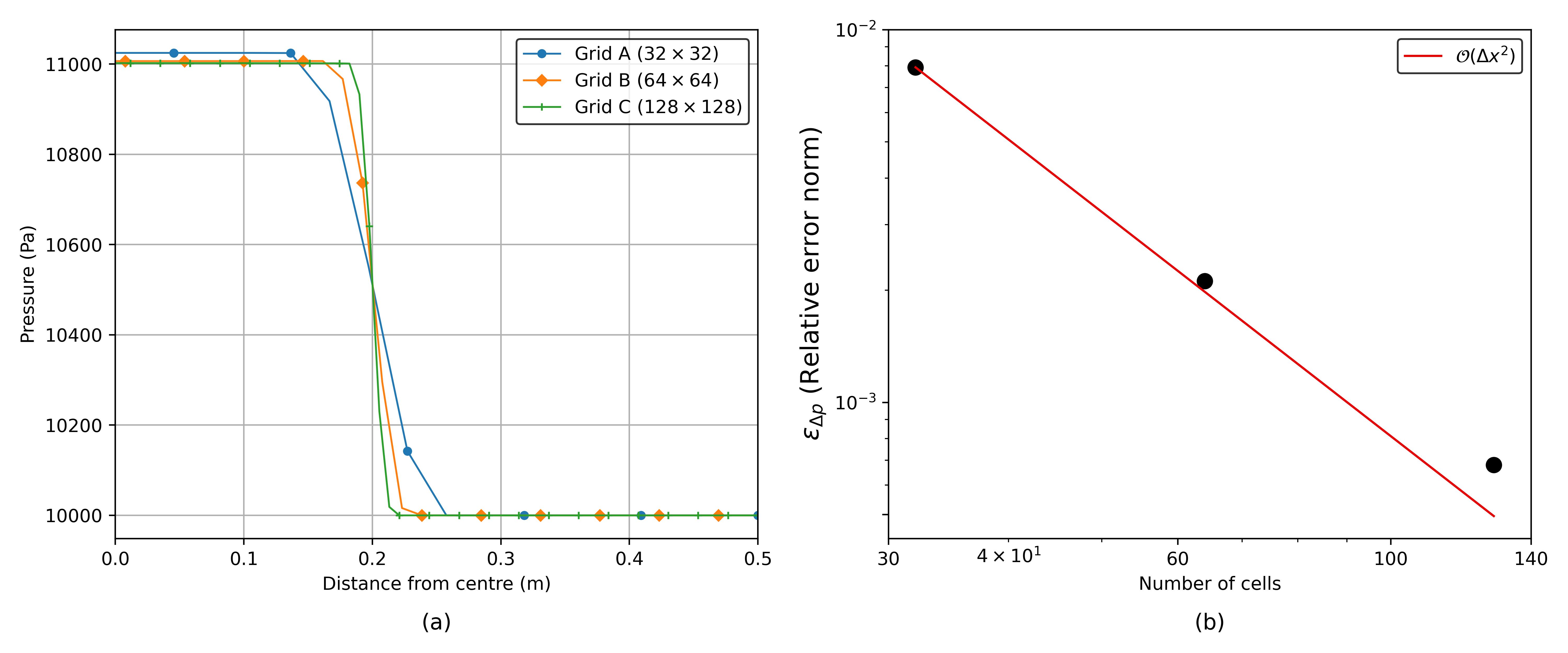}
	\caption{Numerical results for different gird sizes: (a) pressure distribution along the horizontal center line, (b) log–log plot of relative error norms versus the number of cells.}
	\label{fig:Laplace}
\end{figure}

Initially, a uniform pressure of $p_0 = 10^4$~Pa is prescribed throughout the domain. Due to surface tension, the pressure inside the droplet increases to $p_0 + \Delta p$. According to the Young–Laplace law, the exact pressure jump should be $\Delta p^{\text{exact}} = \sigma/R$. However, due to numerical errors from the curvature calculation and spatial discretization, a deviation from the exact pressure jump is expected. To evaluate the accuracy of the proposed numerical method, simulations are performed on three different meshes: Grid A $(32 \times 32)$, Grid B $(64 \times 64)$, and Grid C $(128 \times 128)$. The pressure profiles along the horizontal centerline at $t = 2$~s for all test cases are compared in \fig{fig:Laplace}(a), with the corresponding relative error norms are plotted in \fig{fig:Laplace}(b). The $L_2$ error norm is calculated for cells located strictly within the liquid droplet $\left( \alpha_1 \geq 0.99 \right)$ using the following expression: 
\begin{equation}
	\epsilon_{\Delta p} = \frac{1}{\Delta p^{\text{exact}}} \sqrt{ \sum_{i} \left(\Delta p^{\text{numerical}}_i - \Delta p^{\text{exact}} \right)^2 \Omega_i }.
\end{equation}

From the comparison plot (\fig{fig:Laplace}(a)), it can be seen that a uniform pressure inside the droplet is created, and the pressure jump is converging to the exact solution with almost second-order accuracy (\fig{fig:Laplace}(b)). The numerical error also gives rise to spurious velocities, also known as parasitic currents. For all the test results, these currents remain bounded and maximum value is less than $10^{-2}$ m/s. 

Parasitic currents are typically quantified and analyzed in the presence of physical viscosity using appropriate non-dimensional numbers \cite{garrick2017interface, chang2013direct, abadie2015combined}. Assuming equal viscosities, $\mu_1 = \mu_2 = \mu$, the same static droplet case is simulated on a $64 \times 64$ mesh. The simulations are performed for four different Laplace numbers $\left(La = \frac{\rho \sigma D}{\mu^2}\right)$: $10^4$, $10^6$, $10^9$, and $10^{11}$. A higher Laplace number corresponds to less physical dissipation and, consequently, greater computational difficulty.

\begin{figure}[H]
	\centering
	\includegraphics[width=\textwidth]{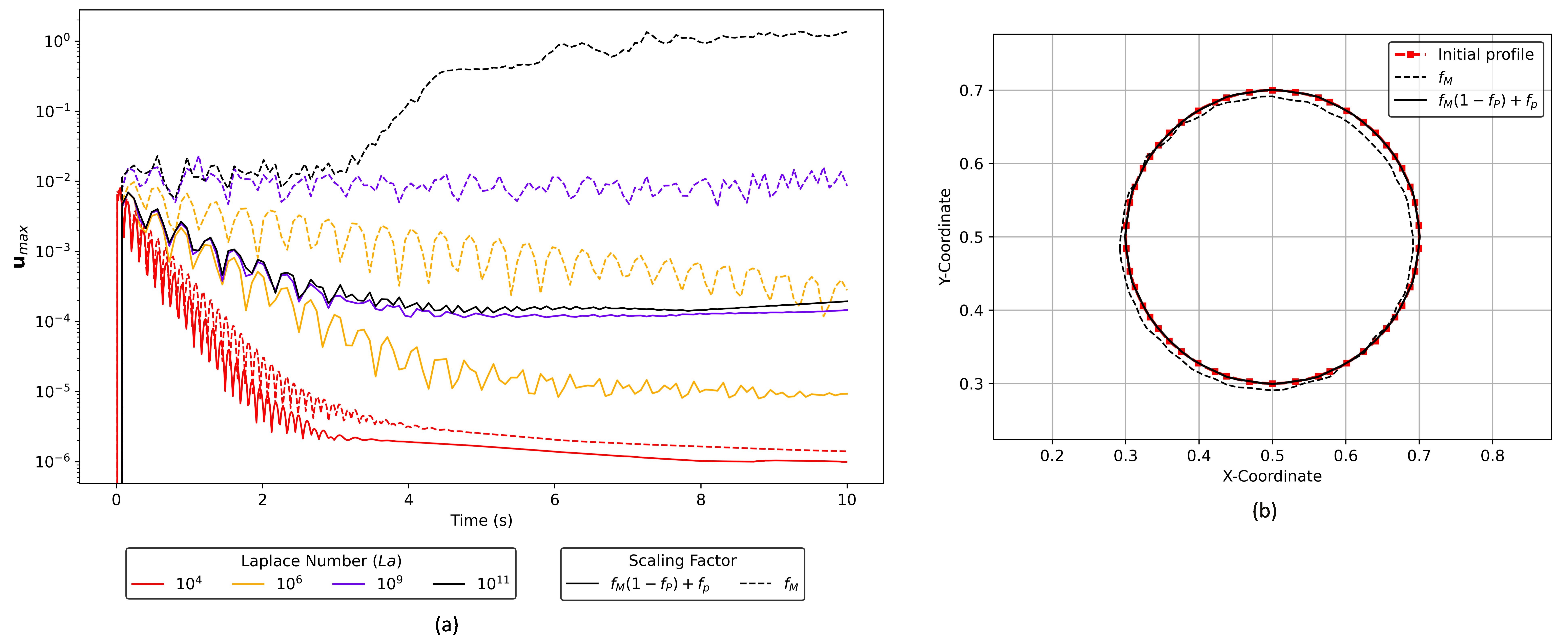}
	\caption{(a) Time history of parasitic currents in a static droplet case, simulated using the simple scaling factor, $f_M$, and the modified scaling factor, $f_M(1 - f_p) + f_p$, for different Laplace numbers. (b) Comparison of the interface profiles for a Laplace number of $10^{11}$ at $t = 9.3$~s, obtained using different scaling factors, against the initial profile.}
	\label{fig:parasitic}
\end{figure}

To demonstrate the effectiveness of the modified scaling factor, numerical results obtained using the simple scaling factor, $f_M$ are compared with those using the modified scaling factor $f_M(1 - fp) + f_p$ in \fig{fig:parasitic}. The time history of the maximum value of the parasitic currents (\fig{fig:parasitic} (a)) reveals that, regardless of the Laplace number, the parasitic current is consistently lower when the modified scaling factor is applied. This reduction occurs because the pressure sensor near the interface diminishes the effect of the low-Mach correction, which subsequently increases numerical dissipation and helps suppress parasitic currents. However, the difference between the parasitic currents observed in the test cases with two different scaling factors decreases at lower Laplace numbers. In case of highest Laplace number, $10^{11}$, the parasitic currents become unbounded if a simple scaling factor is used. These strong currents cause distortion and displacement of the interface, as observed in the comparison of interface profiles at $t = 9.3$~s in \fig{fig:parasitic}(b). Since the low-Mach correction is unavoidable in low-Mach flow regimes, the modified scaling factor offers a balanced approach for achieving stable and accurate solutions.

\subsection{Oscillating droplet}
\begin{figure}[H]
 	\centering
 	\includegraphics[scale = 0.5]{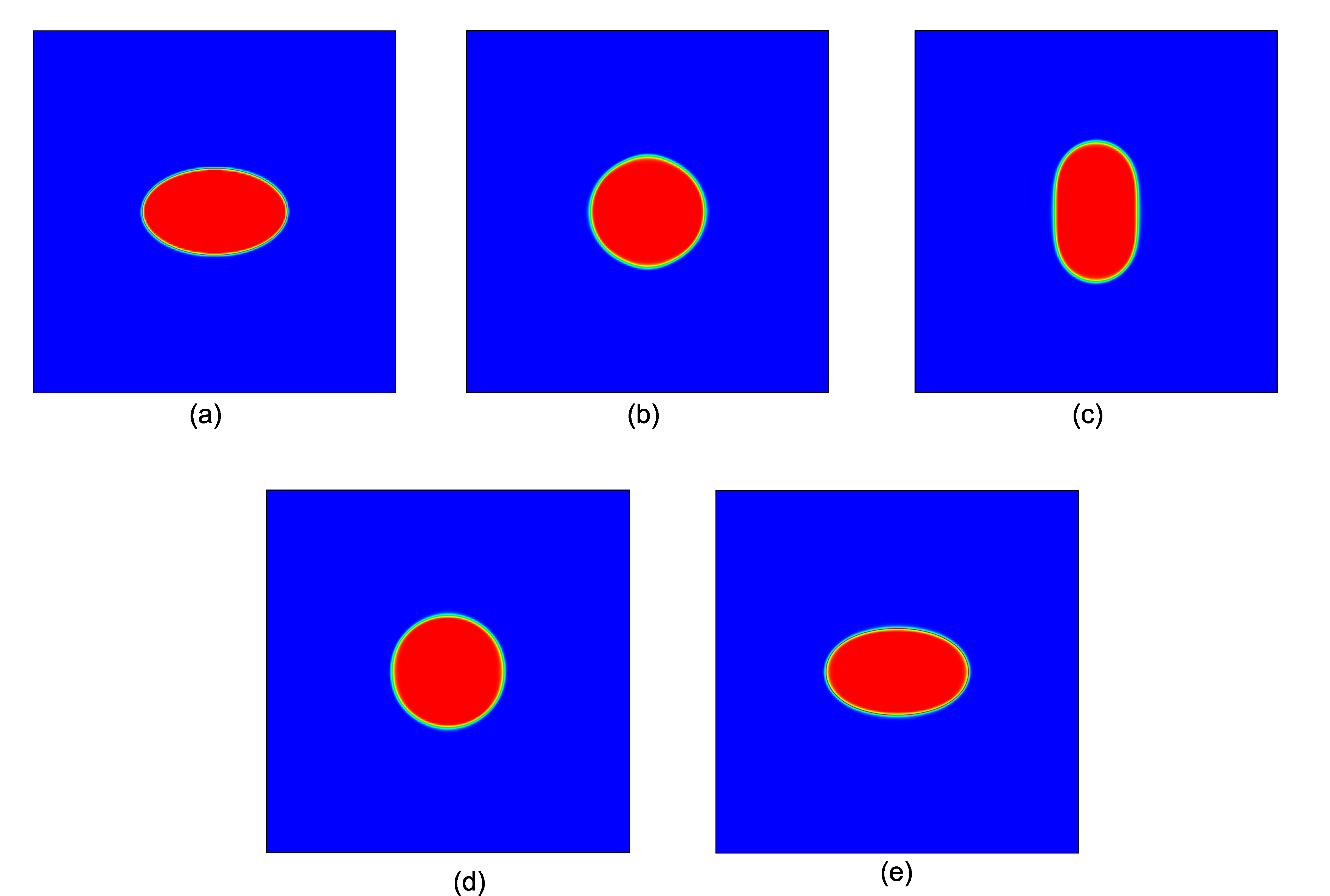}
\caption{Evolution of an oscillating droplet on Grid C ($192 \times 192$) at various normalized times: (a) $t/T = 0$, (b) $t/T = 0.25$, (c) $t/T = 0.5$, (d) $t/T = 0.75$, and (e) $t/T = 1$, where $t$ is the oscillation period.}
 	\label{fig:Oscill_fig}
 \end{figure}

To evaluate the scheme's ability to resolve dynamic, surface tension driven flows, we simulate the standard oscillating droplet benchmark \cite{fiveeq_st_css_2005,seveneq_st_path_2015,panchal2023seven}. In this test case, an ellipsoidal liquid droplet is placed inside a 2D square domain of size $L = 1~\text{m}$, surrounded by air. The initial setup and material properties are adopted from the literature \cite{fiveeq_st_css_2005, seveneq_st_path_2015}. For the liquid (phase 1) and air (phase 2) the properties are defined as
\begin{equation}
	\begin{split}
		 &\rho_1=100~\text{kg/m}^3, \quad \gamma_1=2.4, \quad \text{and}~\pi_1=10^7~\text{Pa}, \\
		 &\rho_2=1~\text{kg/m}^3, \quad \gamma_2=1.4 \quad \text{and}~\pi_2=0~\text{Pa}.
	\end{split}
\end{equation}
Additionally, the surface tension coefficient is set to $\sigma = 342~\text{N/m}$. The initial geometric profile of the liquid droplet is given by the following equation. 
\begin{equation}
	\frac{(x - 0.5)^2}{a^2} + \frac{(y - 0.5)^2}{b^2} = 1. 
\end{equation}
The lengths of the semi-major $(a)$ and semi-minor $(b)$ axes are taken as $a = 0.2~$m and $b = 0.12~$m. For this test problem, the curvature correction formulae (Eq.~\eqref{eq:curv_corr}) is used with two iterations. Due to the non-uniform curvature of the initial liquid-gas interface, the droplet does not remain in its initial shape and undergoes repeated cycles of deformation. The evolution of the droplet is plotted in \fig{fig:Oscill_fig} at different stages of a complete cycle. The oscillation frequency of the droplet is given by the following analytical expression~\cite{rayleigh1879capillary, fyfe1988surface}       
\begin{equation}
	\omega^2 =  \frac{6 \sigma}{(\rho_1 + \rho_2) R^3}. 
\end{equation}  
Here, $R$ represents the equivalent circular radius of the droplet, which is evaluated as $R = \sqrt{ab}$. For an equivalent radius of 0.1549 m, the oscillation period $\left(T = \frac{2 \pi}{\omega}\right)$ evaluates to 0.085 s \cite{panchal2023seven}. 

\begin{figure}[H]
	\centering
	\includegraphics[ width =0.75\textwidth]{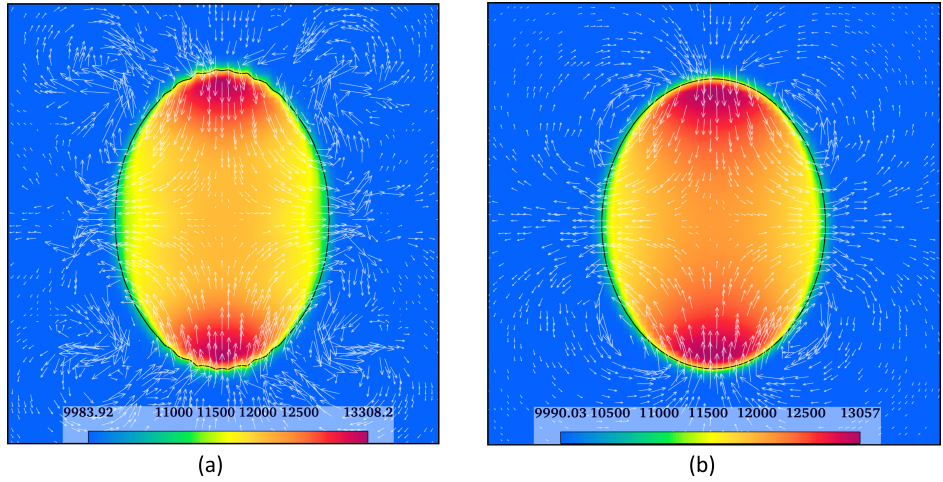}
	\caption{Pressure field (Pa), velocity glyphs, and interface location $(\alpha = 0.5$, black line$)$ at $t = 0.235$ ms for mesh size $128\times128$. (a) Result obtained with scaling factor $f_{\text{lm}} = f_M$. (b) Result obtained with a low Mach correction factor which includes pressure function $(f_{\text{lm}} = f_M(1 - f_p) + f_p)$.}
	\label{fig:Oscill_SenorComp}
\end{figure}

\begin{figure}[H]
   	\centering
   	\includegraphics[width =1\textwidth]{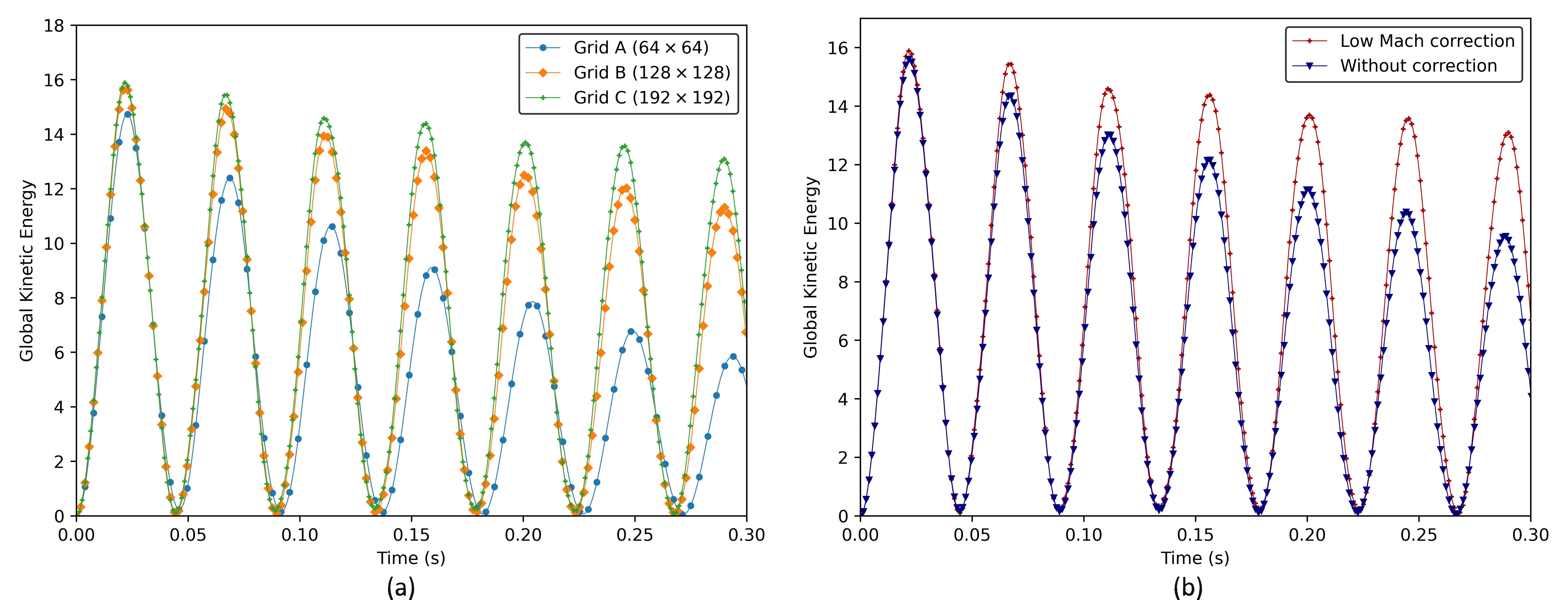}
\caption{Time history of global kinetic energy: (a) comparison of results obtained with the low Mach correction for different mesh sizes, and (b) comparison of results with and without the low-Mach correction on Grid C ($192 \times 192$).}
   	\label{fig:Oscill_plot}
   \end{figure}

Since the pressure jump induced by surface tension is very large, the low Mach correction with pressure function, $(f_{\text{lm}} = f_M(1 - f_p) + f_p)$, is employed for all test cases to suppress spurious interface oscillations. To demonstrate the effect of the pressure function based low Mach correction, the combined pressure contour, velocity vectors, and interface location at $t = 0.235\,\text{ms}$ are presented in \fig{fig:Oscill_SenorComp}. The left plot, obtained using the simple low Mach correction, $f_{\text{lm}} = f_M$, exhibits spurious interface oscillations near the upper and lower ends of the interface, where the pressure gradient is very high. In contrast, when the pressure function based low Mach correction, $f_{\text{lm}} = f_M(1 - f_p) + f_p$, is employed, these spurious interface oscillations disappear, resulting in a smooth interface profile.
During the deformation cycles, kinetic and potential energy are continuously transformed into one another. The temporal variation of the global kinetic energy, defined as $\frac{1}{2}\sum_{i} \rho_i \left| \mathbf{u}_i \right|^2 \Omega_i$, is plotted for different cases in \fig{fig:Oscill_plot}. The kinetic energy reaches its peak value twice during a single physical oscillation cycle, corresponding to the instants when the droplet becomes circular. However, driven by this kinetic energy, the deformation continues until the droplet transforms into a vertical or horizontal elliptical shape, a state that corresponds to minimum kinetic energy. 

Ideally, in the absence of physical viscosity, the cycles of droplet deformation would continue indefinitely. However, inherent numerical dissipation artificially damps the oscillations, an effect that is clearly evident in the kinetic energy plots. As shown in \fig{fig:Oscill_plot}~(a), which compares results obtained across different mesh resolutions using the low Mach correction , numerical dissipation noticeably decreases as grid resolution increases. Additionally, comparing the results with and without the low Mach correction on the finest grid, illustrated in \fig{fig:Oscill_plot}~(b), demonstrates that applying the low Mach correction significantly reduces numerical dissipation. The percentage errors in the computed oscillation frequency $(\omega)$ and energy decay ratio $(G)$ for the different test cases are summarized in \tbl{tab:Oscill_error}. The energy decay ratio $(G)$ is defined as the ratio of the third peak to the first peak of the global kinetic energy curve and quantifies the fraction of kinetic energy remaining after one complete oscillation cycle. Compared to work done by \citet{panchal2023seven}, which reported a $66 \%$ error for $200 \times 200$ mesh, the present work achieves a significant lower error of $4.38 \%$ on a comparable $192 \times 192$ mesh. The results confirm the scheme's robust capability to accurately simulate dynamic, surface tension dominated multiphase flows.
  
  \begin{table}[H]
  	\centering
  	\caption{Percentage error in the droplet oscillation frequency $(\omega)$ and energy decay ratio $(G)$ for different test cases.}
  	\label{tab:Oscill_error}
  	\begin{tabular}{ccc}
  		\hline 
  		Case & Error & G \\
  		\hline 
  		Grid A ($64 \times 64$), Without correction  & 8.151 \% & 0.35 \\
  		Grid A ($64 \times 64$), With correction     & 7.211 \% & 0.722   \\
  		Grid B ($128 \times 128$), Without correction  & 5.223 \%  & 0.717 \\
  		Grid B ($128 \times 128$), With correction  & 4.927 \%  &  0.89   \\
  		Grid C ($192 \times 192$), Without correction   & 4.707 \%  & 0.835 \\
  		Grid C ($192 \times 192$), With correction   & 4.37 \%  &  0.918 \\
  		\hline
  	\end{tabular}
  \end{table}

\begin{figure}[H]
	\centering
	\includegraphics[scale = 0.45]{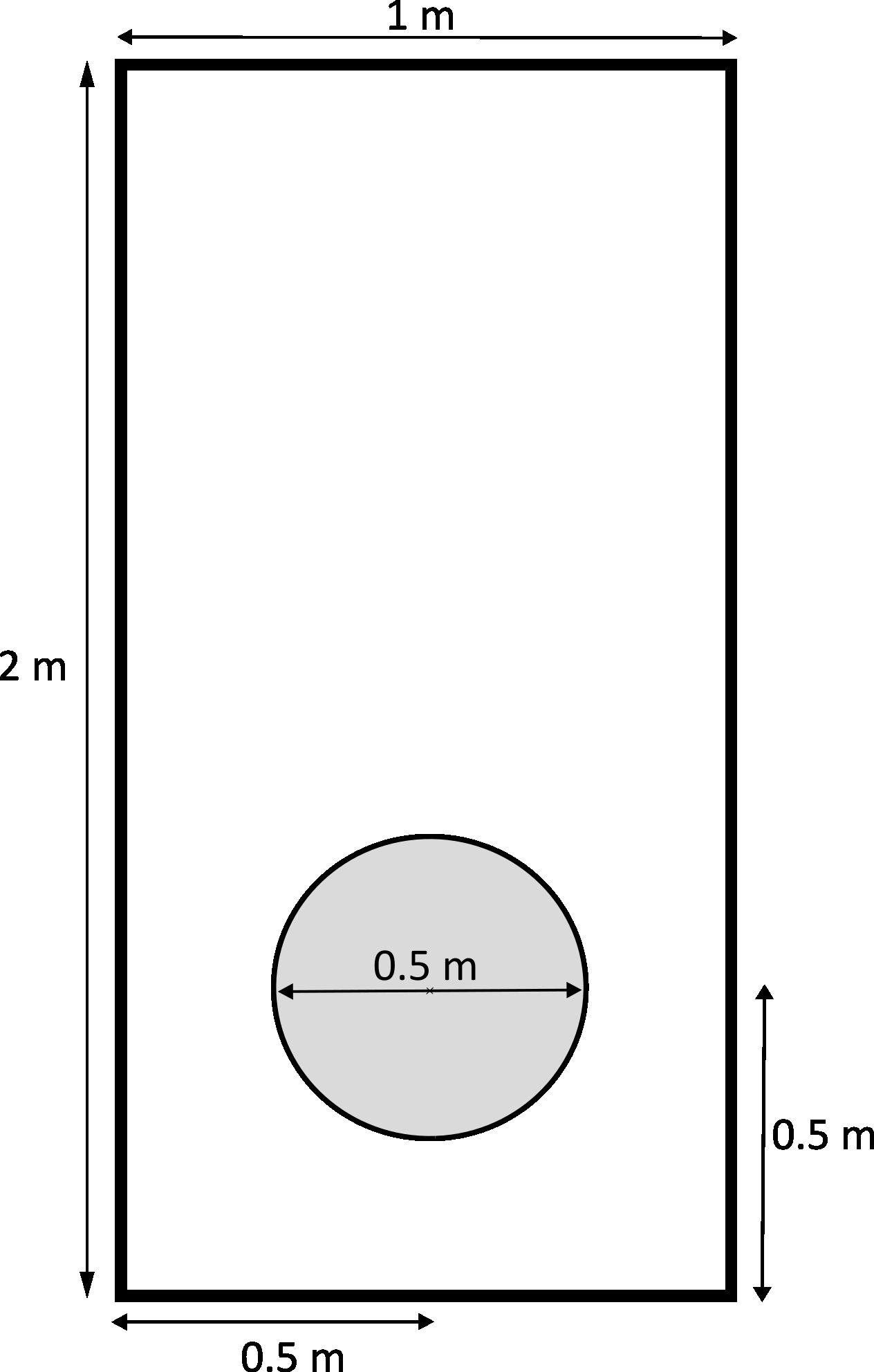}
	\caption{Initial setup for bubble rise test case}
	\label{fig:RB_initial}
\end{figure}

\subsection{Rising bubble}
\begin{figure}[H]
 	\centering
 	\includegraphics[scale = 0.6]{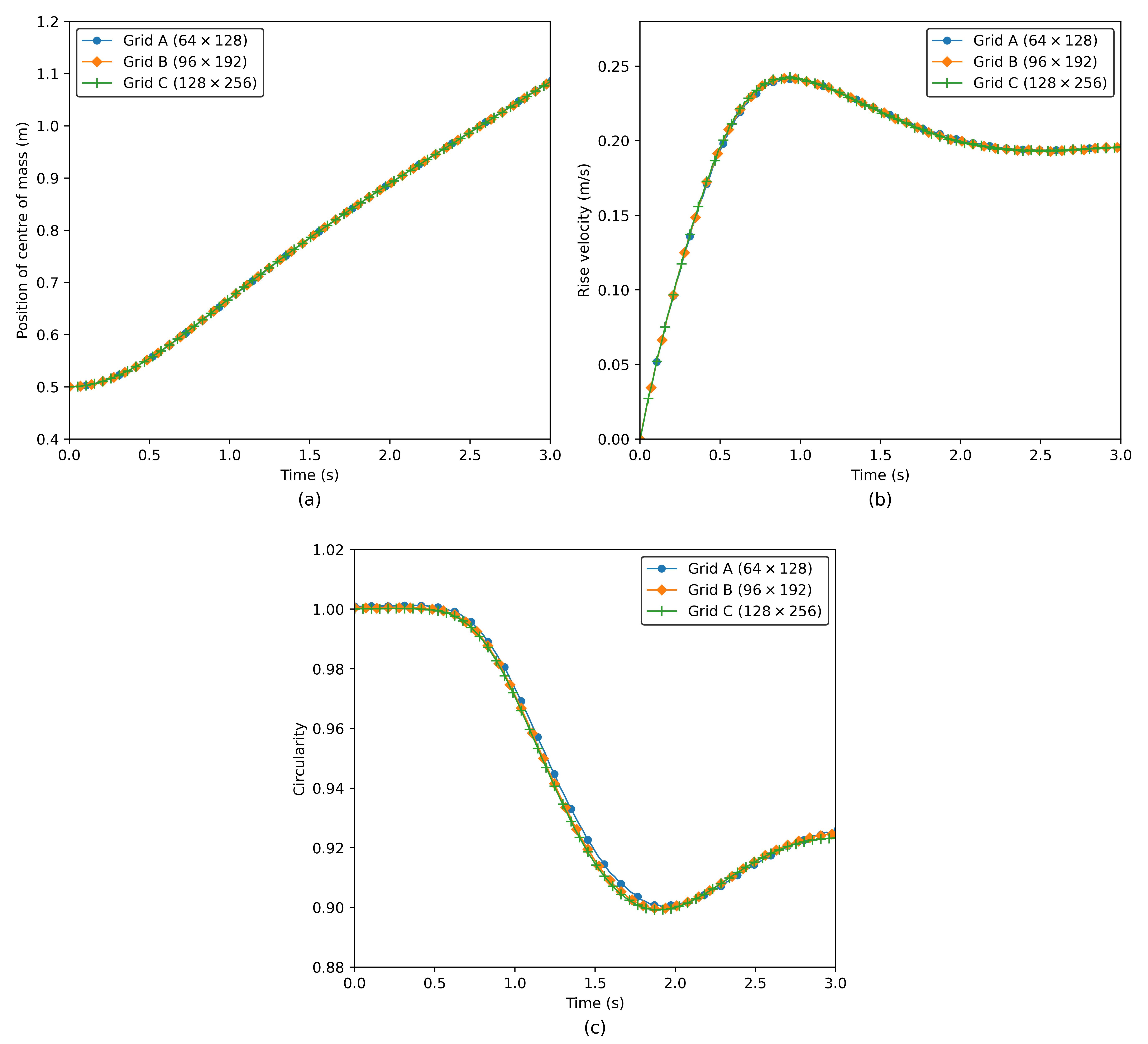}
 	\caption{Comparison of numerical results using the low Mach correction on different meshes: (a) trajectory of the bubble's center of mass $(y_b)$, (b) bubble rising velocity $(v_b)$, and (c) bubble circularity $(\chi_{b})$.}
 	\label{fig:RB_mesh}
 \end{figure}

To test the capability of numerical scheme to accurately resolve low Mach multiphase flows involving gravitation and viscous effect, we consider rising bubble test case \cite{hysing2009quantitative}. The computational setup of the test problem is shown in \fig{fig:RB_initial}. The top and bottom wall considered as no slip wall while free slip boundary condition is imposed on the side walls. The material properties for the bubble (phase 1) and surrounding liquid (phase 2) are taken as 
\begin{equation}
	\begin{split}
		&\rho_1 = 100 ~\text{kg/m}^3, \quad \mu_1 = 1 ~\text{Pa s},  \quad \gamma_1 = 2.4, \quad \text{and}~\pi_1 = 10^7 ~\text{Pa}, \\
		&\rho_2 = 1000 ~\text{kg/m}^3,  \quad \mu_2 = 10 ~\text{Pa s}, \quad \gamma_2 = 4.4, \quad \text{and}~\pi_2 = 6 \times 10^8 ~\text{Pa}. 
	\end{split}
\end{equation}
The simulations are performed using a surface tension coefficient of $\sigma = 24.5~\text{N/m}$ and a gravitational acceleration of $\mathbf{g} = (0, -0.98)~\text{m/s}^2$. The stiffened gas parameters were specifically selected according to the phase densities. The number of iterations used with the curvature correction formula (Eq. \eqref{eq:curv_corr}) is the same as in the previous oscillating droplet problem. The optimal mesh resolution is determined by comparing the time history of bubble's center of mass, its rise velocity and circularity for three different mesh sizes: Grid A $(64 \times 128)$, Grid B $(96 \times 192)$, and Grid C $(128 \times 256)$, as shown in \fig{fig:RB_mesh}. The center of mass, $y_b$, rise velocity, $v_b$, and circularity of the bubble, $\chi_{b}$ are computed as
\begin{equation}
	y_b = \frac{\sum_{i} \left(\alpha_1 \right)_i y_i }{\sum_{i} \left(\alpha_1 \right)_i}, \quad v_b = \frac{\sum_{i} \left(\alpha_1 \right)_i v_i}{\sum_{i} \left(\alpha_1 \right)_i }, \quad \chi_{b} = \frac{P_a}{P_b}. 
\end{equation}
Here, $P_a$ is the perimeter of a circle with area equivalent to bubble and $P_b$ is the actual perimeter of the bubble. Since, the results does not change significantly from Grid B to Grid C, Grid B is selected for further investigation.

\begin{figure}[H]
 	\centering
 	\includegraphics[scale = 0.7]{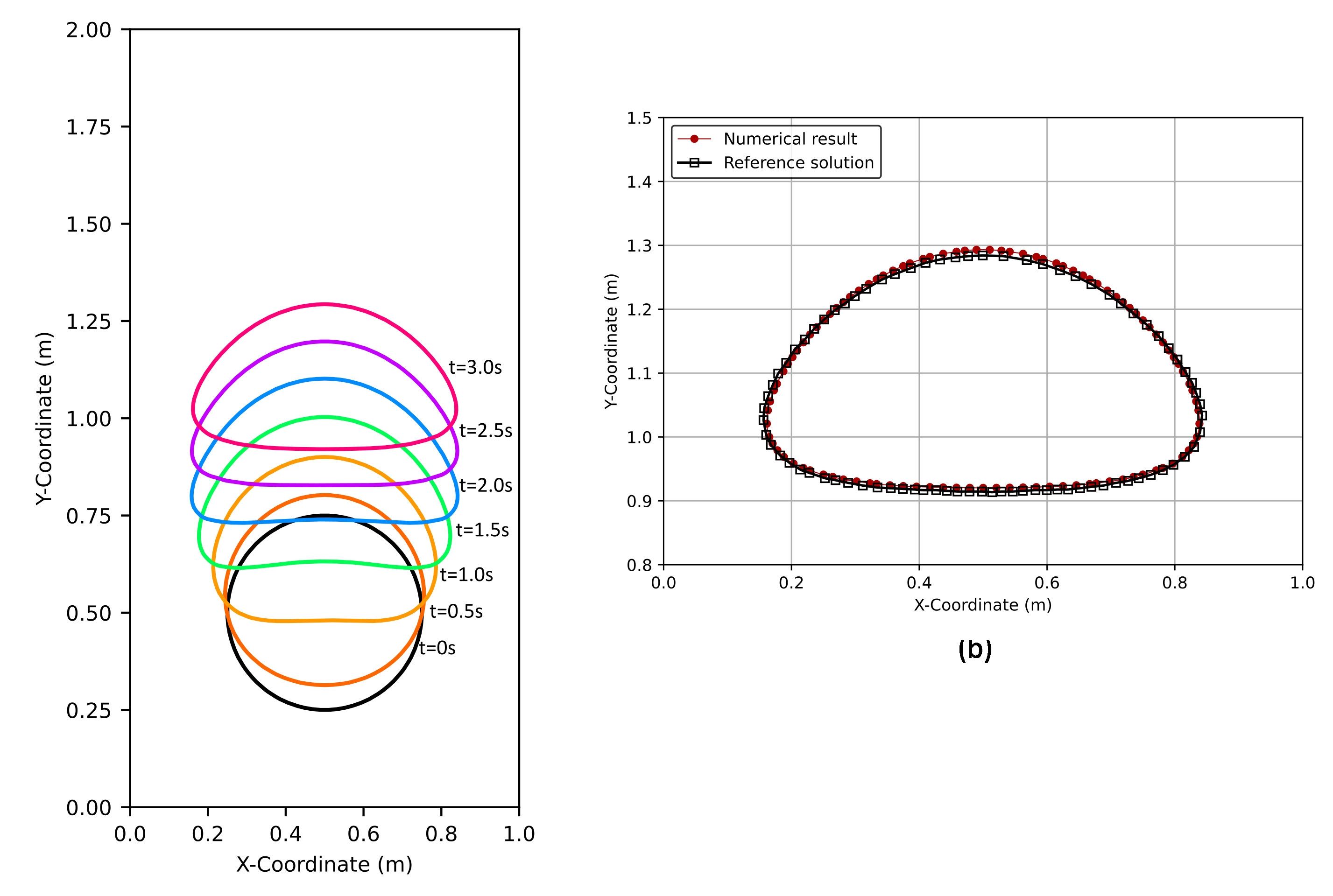}
 	\caption{(a) Temporal evolution of the bubble interface at different time intervals, illustrating movement and deformation. (b) comparison of bubble interface at t = 3~s with the reference incompressible solution \cite{hysing2009quantitative}}
 	\label{fig:RB_interface}
 \end{figure}

\begin{figure}[H]
 	\centering
 	\includegraphics[scale = 0.6]{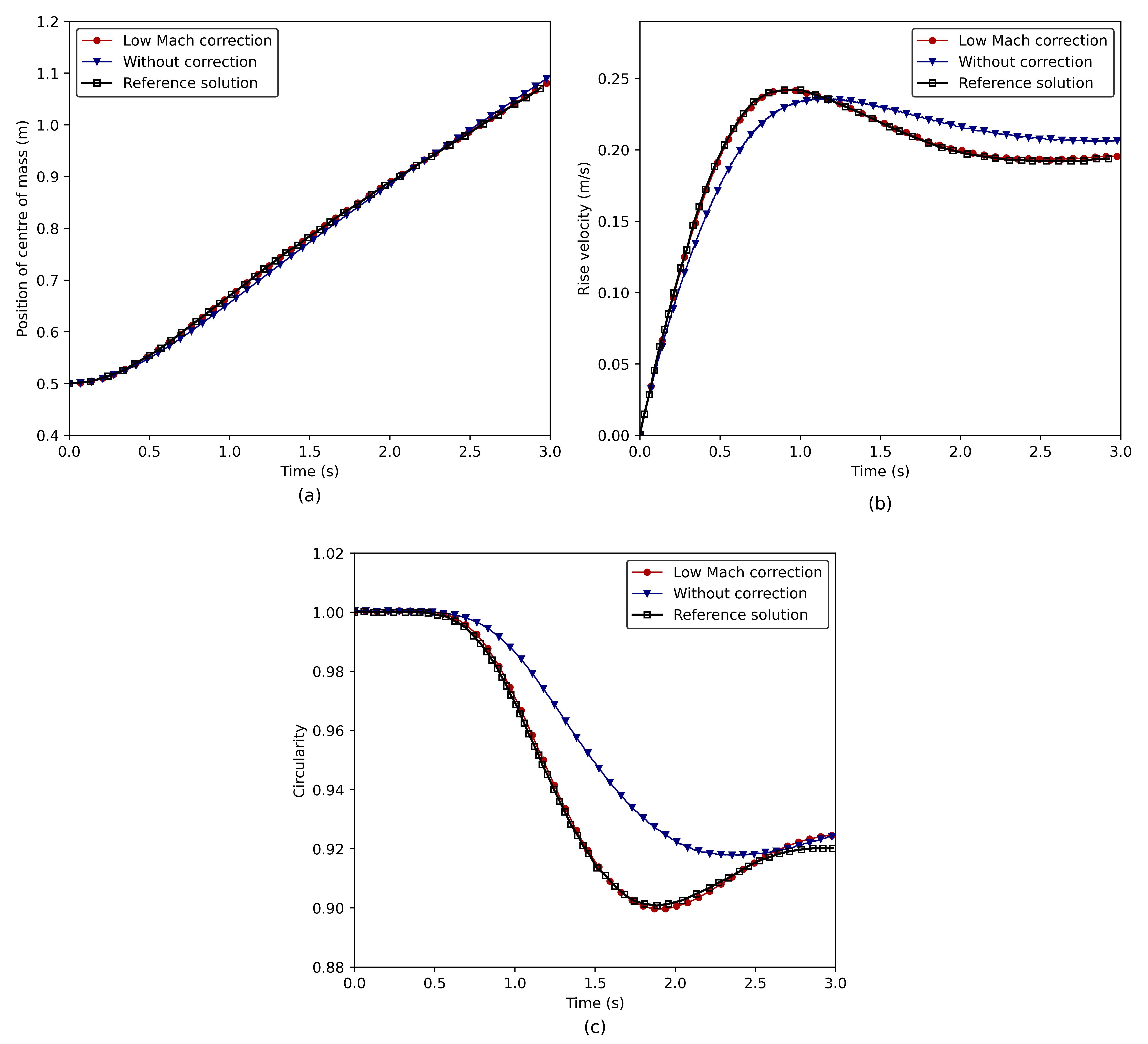}
 	\caption{Comparison of numerical results obtained with and without the low Mach correction against the reference incompressible solution of \citet{hysing2009quantitative}: (a) center of mass trajectory $(y_b)$, (b) rise velocity $(v_b)$, and (c) circularity $(\chi_b)$.}
 	\label{fig:RB_compar}
 \end{figure}

Because of the buoyant force, the bubble starts moving upward (see \fig{fig:RB_interface}(a)), and its velocity increases during the initial phase. As viscous drag increases, the forces balance, and the bubble ultimately attains a constant terminal velocity (see \fig{fig:RB_mesh}~(b)). During its upward motion, the bubble deforms from its initial circular geometry into an ellipsoidal shape, as seen in \fig{fig:RB_interface}~(a). To confirm this behavior, the bubble's shape at 3.0~s is compared against the reference data \cite{hysing2009quantitative} in \fig{fig:RB_interface}~(b), demonstrating close agreement. To illustrate the efficacy of the low Mach correction, numerical results obtained with and without correction are compared against the reference solution \cite{hysing2009quantitative} in \fig{fig:RB_compar}. The comparison clearly indicates that the correction improves the results, which are in close agreement with the incompressible solution. In this case, the low-Mach correction is employed with the modified scaling factor. However, the modified scaling factor is not strictly necessary for this specific test problem, as the physical viscosity alone provides sufficient stability. As demonstrated by the parasitic current study in Section \ref{sec:Static}, the modified scaling factor has a minimal impact at lower Laplace numbers. Because the Laplace number $\left(La = \frac{\rho \sigma D}{\mu^2}\right)$ of the bubble in this setup is relatively low at 1225, the factor's influence remains minimal.
 
\subsection{Shock-induced bubble collapse}
In order to test the robustness of the proposed numerical method, we simulate a standard benchmark problem in which a shock wave traveling in water at Mach 1.72 causes the collapse of an air bubble. This specific configuration has been investigated in numerous studies \cite{nourgaliev2006adaptive,nourgaliev2007high,terashima2009front,terashima2010front,shukla2014nonlinear} to assess the capability of various numerical methods. Initially, an air bubble of diameter $D$ = 6 mm is placed at the origin $(0, 0)$, with a normal shock wave positioned at $x = -5.4$ mm inside a rectangular domain of size $[-12, 18] \times [-7.5, 7.5]~\text{mm}^2$. The computational domain is discretized using a uniform mesh of size $D/250$, and a CFL number of 0.4 is used. The initial conditions for the water and air are defined as follows:
 
 \begin{equation}
 	\begin{split}
 		\text{Post-shock water: } 	&\rho = 1324.81 ~\text{kg/m}^3, p = 1.92 \times 10^9~\text{Pa}, u = 685.2~\text{m/s}, \gamma = 4.4, \pi = 6\times10^8~\text{Pa}, \\
 		\text{Pre-shock water: } 	&\rho = 1000 ~\text{kg/m}^3,  p = 10^5~\text{Pa}, u = 0~\text{m/s}, \gamma = 4.4, \pi = 6\times10^8~\text{Pa}, \\ 
 		\text{Air cavity: } 	&\rho = 1 ~\text{kg/m}^3,  p = 10^5~\text{Pa}, u = 0~\text{m/s}, \gamma = 1.4, \pi = 0~\text{Pa}.
 	\end{split}
 \end{equation}
 
\begin{figure}[H]
	\centering
	\includegraphics[scale = 0.8]{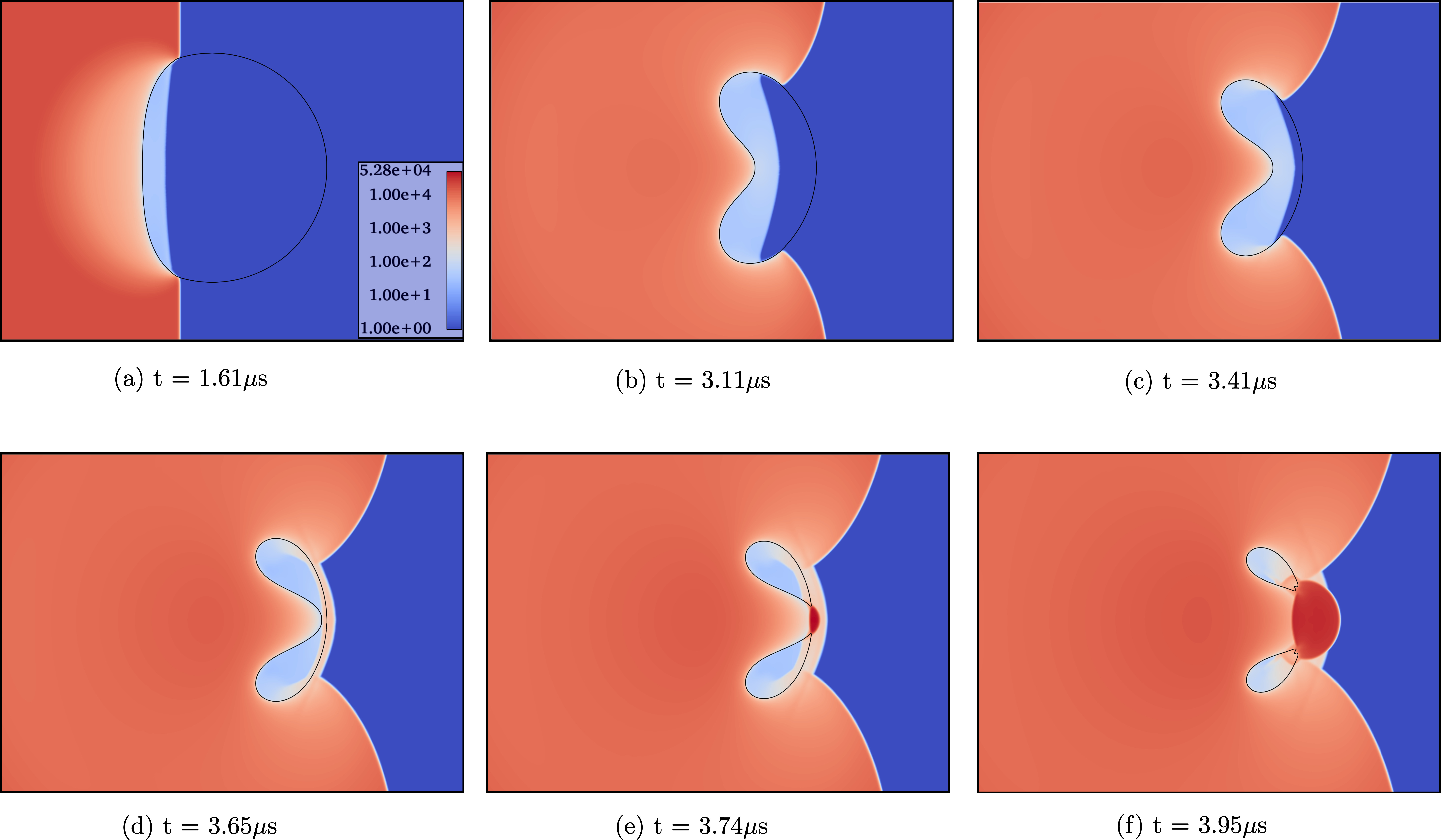}
	\caption{Normalized pressure contours $(p/10^5)$ and interface ($\alpha = 0.5$) at different time instances for shock induced bubble collapse test case.}
	\label{fig:cavity_contour}
\end{figure}

\begin{figure}[H]
	\centering
	\includegraphics[scale = 1.0]{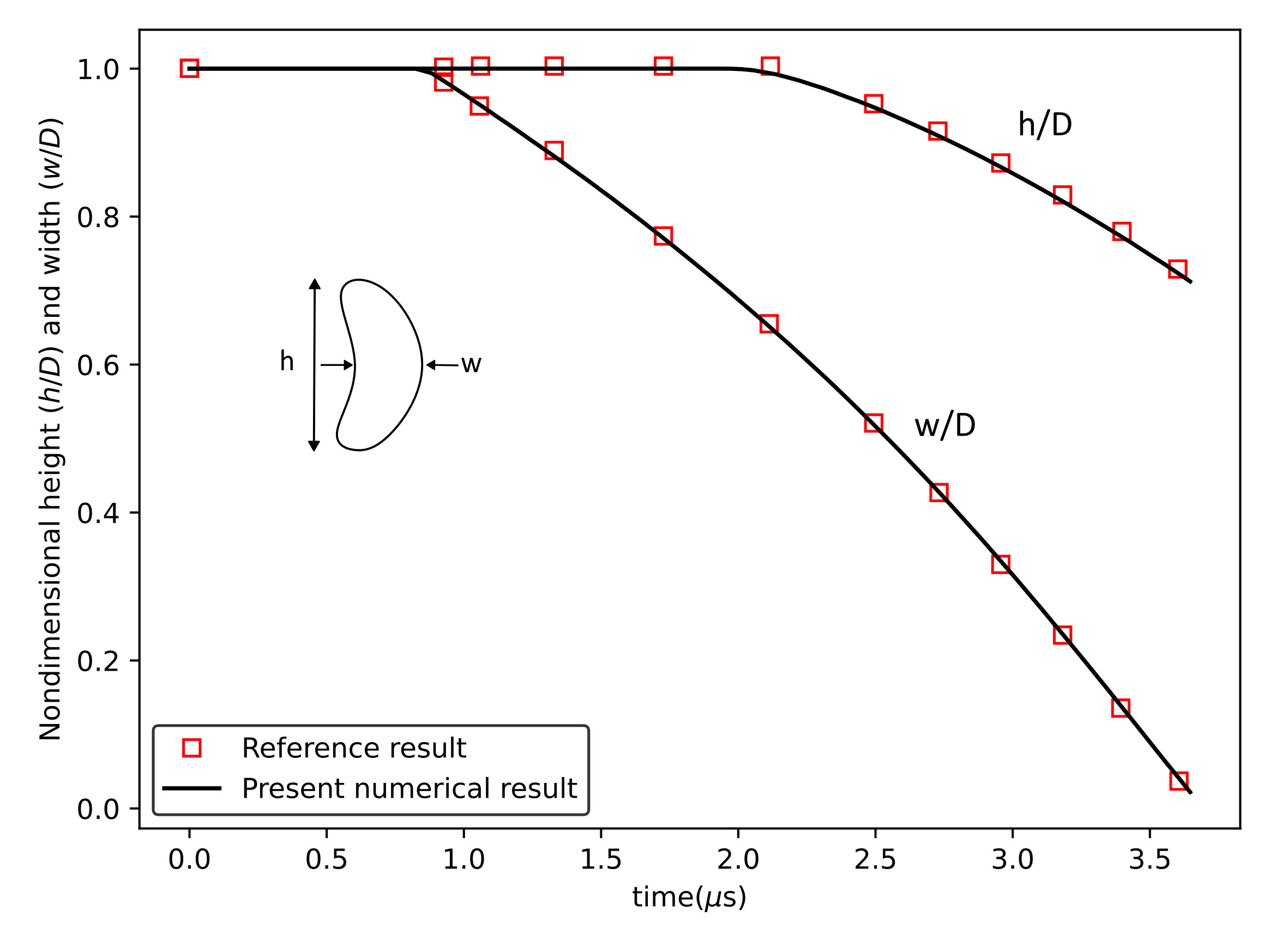}
	\caption{Non-dimensional height (h/D) and width (w/D) of the collapsing bubble and comparison with the reference result from \cite{terashima2010front}}
	\label{fig:cavity_plot}
\end{figure}

Since the Weber number $(We = \frac{\rho u^2 D}{\sigma})$ based on the post-shock conditions is extremely high ($We = 51.83 \times 10^6$), surface tension effects can be safely neglected in this problem. The evolution of the bubble after the collision is presented in \fig{fig:cavity_contour}, which includes a series of normalized pressure contours $(p/10^5)$ and the interface ($\alpha = 0.5$) at various time instants. From \fig{fig:cavity_contour}(a), it can be observed that as the shock wave in the water collides with the bubble interface, a reflected rarefaction in the water and a transmitted shock within the bubble are formed. Following the collision, the bubble begins to deform; specifically, the upstream interface deforms inward while the transmitted shock continues to travel through the cavity. This internal shock reaches the downstream interface before the upstream interface converges upon it. As evident in \fig{fig:cavity_contour}(d), a transmitted shock has already formed in the surrounding water before the bubble breaks apart. When the two sides of the interface meet, the bubble splits in two, and a strong blast wave forms behind the transmitted wave (see \fig{fig:cavity_contour}(e)). For a more detailed explanation of the flow physics, refer the work of \citet{hawker2012interaction}. The numerical results are in qualitative agreement with previous studies \cite{nourgaliev2007high,hawker2012interaction,shukla2014nonlinear}. For quantitative validation, we plotted the nondimensional height and width of the bubble in \fig{fig:cavity_plot} prior to its collapse and compared against the results of \cite{terashima2010front}. The results presented in \fig{fig:cavity_contour} and \fig{fig:cavity_plot} demonstrate that the present numerical scheme successfully resolves high-speed flows with shock waves, even when the involved fluids exhibit vastly different compressibilities. 

In this work, the stiffened gas equation of state (EOS) is used for water, which is known to produce non-physical negative pressures, as widely reported in the literature \cite{saurel1999,saurel2001, chang2007robust,garrick2017interface}. The present method robustly circumvents this issue by incorporating an implicit approximation, $\bar{p}_I={p}^{*}_I$, within the pressure relaxation step. Furthermore, to prevent computational breakdown caused by large negative pressures during the evolution step, the flux computation procedure for this test case relies on the isentropic relation in Eq. \eqref{eq:Prel2} rather than Eq. \eqref{eq:Prel1}. For a detailed explanation, please refer to Section \ref{sec:failure}.

\section{Conclusion}\label{sec:conclusion} 
In this work, we presented an effective finite volume method capable of resolving multiphase flows involving surface tension, viscous, and gravitational effects across all Mach numbers. We incorporated capillary effects into the two-pressure six-equation model using the continuum surface force (CSF) approach \cite{brackbill1992continuum} as a source term. The curvature values at the cell centers are computed using a weighted least squares method on a smoothed field obtained via a convolution process. Furthermore, both the HLLC Riemann solver and the pressure relaxation step were modified to consistently account for capillary-induced pressure jumps, preserving the Laplace pressure discontinuity during phase equilibration. The pressure jump across the contact wave was derived based on a generalised Riemann invariant analysis performed for the six-equation model with capillary terms.  

To accurately compute multiphase flow problems in the low Mach regime while overcoming excessive numerical diffusion, we extended the low Mach correction approach presented in our recent work \cite{bharate2025enhanced}. Although this method significantly improves accuracy without imposing strict time-step constraints, it is prone to instabilities in regions of strong pressure variation. To mitigate this issue, we introduced a modified scaling factor using a multidimensional pressure-based sensor, providing a stable and robust approach across low Mach regimes.  Demonstrated across the test cases, this pressure sensor-based formulation successfully eliminates spurious oscillations that occur under simple low Mach corrections near regions of high-pressure gradients, yielding smooth interface profiles.

The efficacy of the proposed numerical framework was validated against several benchmark problems involving various physical phenomena. For the oscillating droplet case, key quantitative metrics including the percentage errors in computed oscillation frequency ($\omega$) and energy decay ratio (G) - representing the remaining kinetic energy fraction after one complete cycle - demonstrated strong accuracy, while the rising bubble results confirmed the scheme’s capability to capture interface dynamics and low Mach flow features. Furthermore, the parasitic current study involving a static droplet highlights the role of the modified scaling factor in dampening unwanted numerical currents and maintaining solution stability, particularly for flows with lower physical viscosity. The effectiveness of this scaling factor is further supported by the dam break problem, which demonstrated improved agreement with experimental data. Finally, the robustness of the overall scheme in handling highly compressible regimes and preventing negative pressures was confirmed through the simulation of a shock-induced bubble collapse.

\appendix

\section{Generalised Riemann Invariants Analysis} \label{sec:GRIs}

The approximate Riemann solver relies on relations across the waves, which are determined using Generalised Riemann Invariants (GRI) analysis \cite{riemann1997}. To incorporate capillary effects into the solver, a constant curvature is assumed \cite{fiveeq_st_css_2005, fiveeq_st_csf_2017}. Considering only surface tension effects, the multiphase system in Eq.~\eqref{eq:model}, can be expressed in an augmented one-dimensional form as:
\begin{equation}
	\begin{gathered}
	\frac{\partial \hat{\mathbf{U}}}{\partial t} + \frac{\partial \mathbf{F}^{c} ( \hat{\mathbf{U}} ) }{\partial \hat{x}} + \mathsf{B}  ( \hat{\mathbf{U}} ) \frac{\partial  \hat{\mathbf{U}}}{\partial \hat{x}} = 0 \\
	\hat{\mathbf{U}} = \left[\begin{array}{c} \alpha_1 \\ \alpha_1 \rho_1 \\ \alpha_2 \rho_2 \\  \rho u_n \\ \rho u_t \\ \alpha_1 \rho_1 e_1 \\ \alpha_2 \rho_2 e_2 \end{array}\right], \quad  \mathbf{F}^{c} ( \hat{\mathbf{U}} )  = \left[\begin{array}{c} \alpha_1 u_n \\ \alpha_1 \rho_1 u_n \\  \alpha_2 \rho_2 u_n  \\  \rho u^2_n +  p  \\ \rho u_n u_t \\ \alpha_1 \rho_1 e_1 u_n \\ \alpha_2 \rho_2 e_2 u_n \end{array}\right], \quad \mathsf{B}  ( \hat{\mathbf{U}} ) = \left[ \begin{matrix}
	0  & \frac{\alpha_1 u_n}{\rho} & \frac{\alpha_1 u_n}{\rho} & -\frac{\alpha_1}{\rho} & 0 & 0 & 0 \\
    0  & 0 & 0 & 0 & 0 & 0 & 0 \\
    0  & 0 & 0 & 0 & 0 & 0 & 0 \\
	\sigma \kappa  & 0 & 0 & 0 & 0 & 0 & 0 \\
	0  & 0 & 0 & 0 & 0 & 0 & 0 \\
	0  & -\frac{\alpha_1 p_1 u_n}{\rho} & -\frac{\alpha_1 p_1 u_n}{\rho} & \frac{\alpha_1 p_1}{\rho} & 0 & 0 & 0 \\
	0  & -\frac{\alpha_2 p_2 u_n}{\rho} & -\frac{\alpha_2 p_2 u_n}{\rho} & \frac{\alpha_2 p_2 }{\rho} & 0 & 0 & 0 
	\end{matrix}  \right]
\end{gathered}
\end{equation}
Above system can be written in quasi-linear form as  
\begin{equation} 
\frac{\partial \mathbf{\hat{W}}}{\partial t} + \mathsf{A} (\mathbf{\hat{W}}) \frac{\partial \mathbf{\hat{W}}}{\partial \hat{x}}  = 0. 
\end{equation}
Here, $\mathbf{\hat{W}}$ is the vector with primitive variables and coefficient matrix $\mathsf{A} (\mathbf{\hat{W}})$ is obtained using the relation 
\begin{equation}
	\mathsf{A}(\mathbf{\hat{W}}) =  \frac{\partial \hat{\mathbf{W}}}{\partial \hat{\mathbf{U}}} \left( \frac{\partial \mathbf{F}^{c} ( \hat{\mathbf{U}} ) }{\partial \hat{\mathbf{U}}} + \mathsf{B}  ( \hat{\mathbf{U}} )  \right) \frac{\partial \hat{\mathbf{U}}}{\partial \hat{\mathbf{W}}}
\end{equation}
where $\mathbf{\hat{W}}$ and $\mathsf{A}(\mathbf{\hat{W}})$ are given below
\begin{equation}
\mathbf{\hat{W}} =  \left[ \begin{array}{c}
    \alpha_1 \\ \rho_1 \\ \rho_2 \\ u_n \\ u_t \\ p_1 \\ p_2
\end{array} \right]  , \quad \mathsf{A}(\mathbf{\hat{W}}) = \left[ \begin{matrix}
        u_n & 0 & 0 & 0 & 0 & 0 & 0\\
        0 & u_n & 0 & \rho_1 & 0 & 0 & 0 \\
        0 & 0 & u_n & \rho_2  & 0& 0 & 0 \\
        \noalign{\medskip} \dfrac{p_1 - p_2 - \sigma \kappa}{\rho} & 0 & 0 & u_n & 0 & \dfrac{\alpha_1}{\rho} & \dfrac{\alpha_2}{\rho} \\
        0 & 0 & 0 & 0 & u_n & 0 & 0 \\
        0 & 0 & 0 & \rho_1 a^2_1 & 0 & u_n & 0 \\
        0 & 0 & 0 & \rho_2 a^2_2 & 0 & 0 & u_n \\
    \end{matrix}  \right].
\end{equation}
Matrix $ \mathsf{A}(\mathbf{\hat{W}})$ has seven eigenvalues as 
\begin{equation}
\lambda_1 = u_n - a, \quad \lambda_{2,..,6} = u_n, \quad \lambda_7 = u_n + a.
\end{equation}
and the corresponding right eigenvectors are  
\begin{equation}\label{eq:eigenvectos}    
    \begin{gathered}
        \mathbf{R}^1 = \left[ \begin {array}{c} 0\\ \noalign{\medskip}{\dfrac {\rho_{{1}}}{
\rho_{{2}}{a_{{2}}}^{2}}}\\ \noalign{\medskip}\dfrac{1}{a_{2}^{2}}
\\ \noalign{\medskip}-{\dfrac {a}{\rho_{{2}}{a_{{2}}}^{2}}}
\\ \noalign{\medskip}0\\ \noalign{\medskip}{\dfrac {\rho_{{1}}{a_{{1}}}
^{2}}{\rho_{{2}}{a_{{2}}}^{2}}}\\ \noalign{\medskip}1\end {array}
 \right], \quad
          \mathbf{R}^2 = \left[ \begin {array}{c} 0\\ \noalign{\medskip}1\\ \noalign{\medskip}0
\\ \noalign{\medskip}0\\ \noalign{\medskip}0\\ \noalign{\medskip}0
\\ \noalign{\medskip}0\end {array} \right] , \quad 
\mathbf{R}^3 = \left[ \begin {array}{c} {\dfrac {-\alpha_{{2}}}{
p_{{1}}-p_{{2}}- \sigma\,\kappa}}\\ \noalign{\medskip}0\\ \noalign{\medskip}0\\ \noalign{\medskip}0
\\ \noalign{\medskip}0\\ \noalign{\medskip}0\\ \noalign{\medskip}1
\end {array} \right] \\
 \mathbf{R}^4 =  \left[ \begin {array}{c} {\dfrac {-\alpha_{{1}}}{p_{{1}}-p_{{2}} -\sigma\,\kappa }}\\ \noalign{\medskip}0\\ \noalign{\medskip}0\\ \noalign{\medskip}0
\\ \noalign{\medskip}0\\ \noalign{\medskip}1\\ \noalign{\medskip}0
\end {array} \right] , \quad
 \mathbf{R}^5 = \left[ \begin {array}{c} 0\\ \noalign{\medskip}0\\ \noalign{\medskip}0
\\ \noalign{\medskip}0\\ \noalign{\medskip}1\\ \noalign{\medskip}0
\\ \noalign{\medskip}0\end {array} \right] , \quad
 \mathbf{R}^6 =  \left[ \begin {array}{c} 0\\ \noalign{\medskip}0\\ \noalign{\medskip}
1\\ \noalign{\medskip}0\\ \noalign{\medskip}0\\ \noalign{\medskip}0
\\ \noalign{\medskip}0\end {array} \right] , \quad
\mathbf{R}^7 =  \left[ \begin {array}{c} 0\\ \noalign{\medskip}{\dfrac {\rho_{{1}}}{
\rho_{{2}}{a_{{2}}}^{2}}}\\ \noalign{\medskip}\dfrac{1}{a_{2}^{2}}
\\ \noalign{\medskip}{\dfrac {a}{\rho_{{2}}{a_{{2}}}^{2}}}
\\ \noalign{\medskip}0\\ \noalign{\medskip}{\dfrac {\rho_{{1}}{a_{{1}}}
^{2}}{\rho_{{2}}{a_{{2}}}^{2}}}\\ \noalign{\medskip}1\end {array}
 \right]. 
     \end{gathered}
 \end{equation}
From the above expressions, it is evident that the eigenvalues of the multiphase system do not change even after the inclusion of the capillary term. In these expressions, $a$ denotes the mixture sound speed, which can be computed as
\begin{equation}\label{eq:soundspeed}
	a = \sqrt{\frac{1}{\rho} \sum_{j} \alpha_j \rho_j a^2_j}.
\end{equation}
 
With right eigenvector $\mathbf{R}^i = {\left[R^i_1, R^i_2,.., R^i_7\right]}^T$, the GRIs across wave $\lambda_i$ in the Riemann solution can be found using following relation \cite{riemann1997}
\begin{equation}\label{eq:}
\frac{d \alpha_1}{R^i_1} = \frac{d \rho_1}{R^i_2} = \frac{d \rho_2}{R^i_3} = \frac{d u_n}{R^i_4} = \frac{d u_t}{R^i_5} = \frac{d p_1}{R^i_6} = \frac{d p_2}{R^i_7}
\end{equation}
While applying above expression for eigenvectors \eqref{eq:eigenvectos} we may encounter zeros in some of the denominator terms. This implies that the change in corresponding variable (appearing in the numerator) is zero. Using GRIs analysis on $\mathbf{R}^1$, we obtain following important relations across $\lambda_1$ wave
\begin{equation}\label{eq:left_wave}
	d\alpha_1 = 0, \quad du_t = 0.
\end{equation}
Similarly for $\lambda_7$ wave we have 
\begin{equation}\label{eq:right_wave}
	d\alpha_1 = 0, \quad du_t = 0.
\end{equation}

In the 6-equation compressible multiphase framework, the intermediate eigenvalues $\lambda_{2}, \dots, \lambda_{6}$ are represented by a single contact speed $S^*$, effectively merging their corresponding characteristic fields into a single middle wave within the HLLC Riemann solver structure. The invariants across this discontinuity can be found by performing GRI analysis on all the right eigenvectors associated with eigenvalues $\lambda_{2,..,6} = u$. Using $\mathbf{R}^3$ we get the following relations,
\begin{equation}
d\alpha_1 (p_1 - p_2 -\sigma \kappa ) = dp_2 (-\alpha_2), \quad d\rho_1 = 0, \quad d\rho_2 = 0, \quad du_n = 0, \quad du_t = 0, \quad dp_1 = 0
\end{equation}
From the relation  $d\alpha_1 (p_1 - p_2 -\sigma \kappa ) = -dp_2 \alpha_2$ and $dp_1 = 0$, we obtain another expression, $dp = \sigma \kappa d\alpha_1$. Similarly, using $\mathbf{R}^4$ we obtain 
\begin{equation}
d\alpha_1 (p_1 - p_2 -\sigma \kappa ) = dp_1 (-\alpha_1), \quad d\rho_1 = 0, \quad d\rho_2 = 0, \quad du_n = 0, \quad du_t = 0, \quad dp_2 = 0
\end{equation}
From the relation $d\alpha_1 (p_1 - p_2 -\sigma \kappa ) = -dp_1\alpha_1$ and $dp_2 = 0$, we again obtain,  $dp = \sigma \kappa d\alpha_1$. This implies that mixture pressure follows Laplace law across the intermediate wave. 

From the GRI analysis using $\mathbf{R}^3$ and $\mathbf{R}^4$, we also find that the phasic densities $(\rho
_1, \rho_2)$, normal velocity $(u_n)$, and tangential velocity $(u_t)$ remain constant. However, examining other eigenvectors, $\mathbf{R}^1$, $\mathbf{R}^5$ and $\mathbf{R}^6$, we conclude that except for normal velocity $(u_n)$ all other variable may change across the wave. Finally, the following relations holds across the intermediate wave,
\begin{equation}\label{eq:middle_wave}
dp = \sigma \kappa d\alpha_1, \quad du_n = 0
\end{equation} 

\section{Preventing Computational Breakdown in Strong Shock Waves via Isentropic Pressure Relations}\label{sec:failure} 

Phasic pressures computed from the phasic energy equations act as intermediate values for determining the final relaxed pressure. Although they are subsequently corrected using the mixture total energy equation, the phasic pressures from the evolution step can still influence the volume fraction and final relaxed pressure. Thus, to avoid computational breakdown due to negative volume fractions or negative pressures after relaxation, we can use measures to minimize the influence of the phasic pressures from the evolution step in the presence of strong shock waves. Primarily, we must avoid large negative pressures from the evolution step. To achieve this, rather than relying on the expression suggested by \citet{Saurel2009SimpleAE} (Eq.~\eqref{eq:Prel1}), we instead utilize the isentropic relation from Eq.~\eqref{eq:Prel2}.

\begin{figure}[H]
	\centering
	\includegraphics[width=\textwidth]{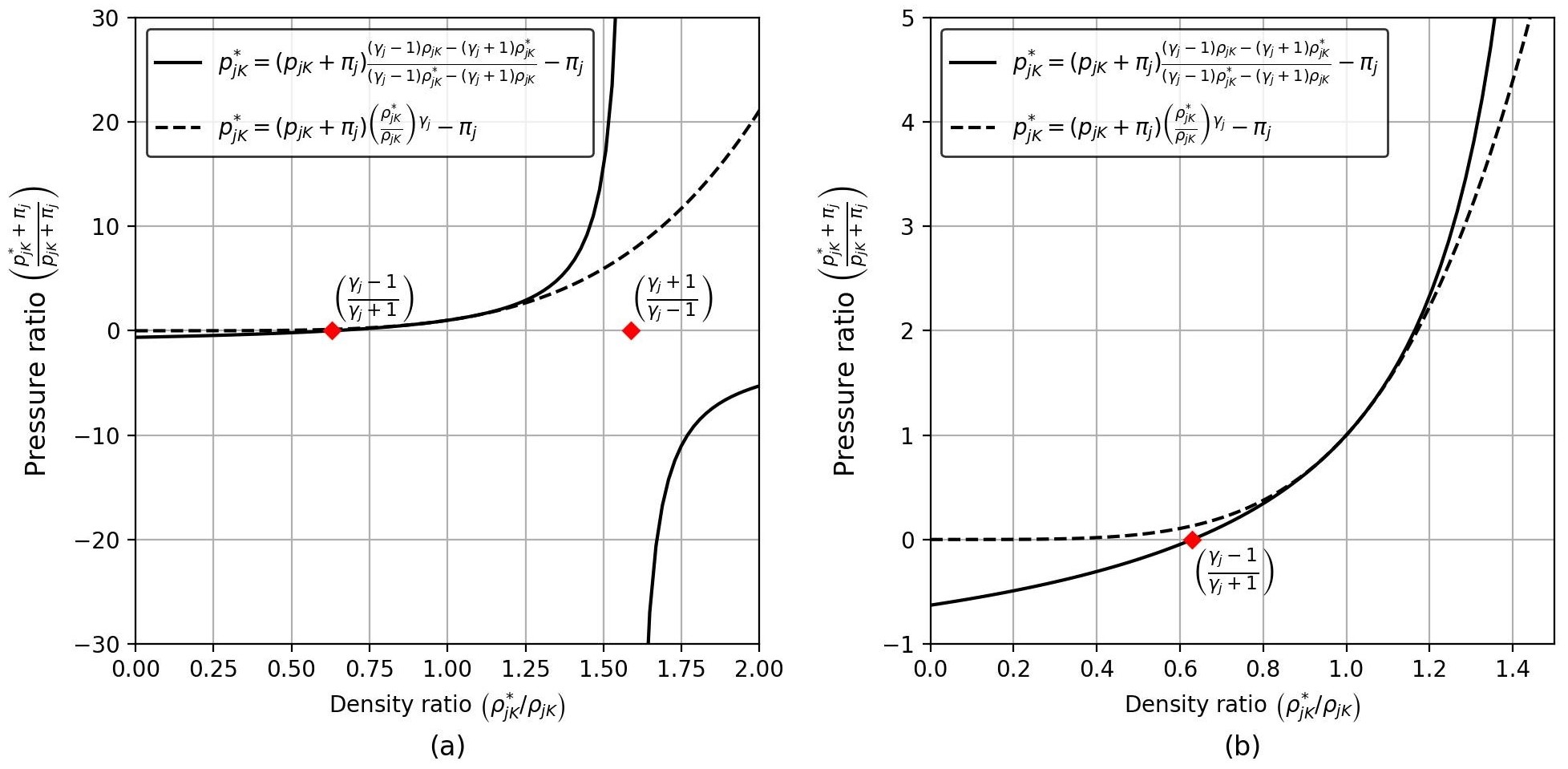}
	\caption{Pressure ratio $\left(\frac{p^{*}_{jK} + \pi_{j}}{p_{jK} + \pi_{j}}\right)$ versus density ratio $\left(\frac{\rho^{*}_{jK}}{\rho_{jK}}\right)$ across the left or right wave, computed via \eqref{eq:Prel1} and \eqref{eq:Prel2} for $\gamma = 4.4$. (a) Broader view, and (b) magnified view.}
	\label{fig:pJump}
\end{figure}

If we plot the pressure ratio $\left(\frac{p^{*}_{jK} + \pi_{j}}{p_{jK} + \pi_{j}}\right)$ using both expressions given in Eqs.~\eqref{eq:Prel1} and \eqref{eq:Prel2} for a typical water value of $\gamma=4.4$ (\fig{fig:pJump}), we observe that for moderate density ratios (near $\frac{\rho^{*}_{jK}}{\rho_{jK}} = 1$), both expressions converge to the same value. Additionally, \fig{fig:pJump} shows that Eq.~\eqref{eq:Prel1} yields a negative pressure ratio for $\left(\frac{\rho^{*}_{jK}}{\rho_{jK}}\right) < \left(\frac{\gamma_j -1}{\gamma_j +1}\right)$ and $\left(\frac{\rho^{*}_{jK}}{\rho_{jK}}\right) > \left(\frac{\gamma_j +1}{\gamma_j -1}\right)$. In contrast, Eq.~\eqref{eq:Prel2} consistently yields a positive pressure ratio. Therefore, utilizing the isentropic relation from Eq.~\eqref{eq:Prel2} effectively prevents large negative pressures during the evolution step.

Another observation regarding problems that involve high-strength shock waves is the influence of the approximate interface pressure, $\bar{p}_I$, during the relaxation procedure. If we use an implicit approximation, such that $\bar{p}_I = {p}^{*}_I$, we can further minimize the influence of large negative pressures arising from the evolution step.

\bibliographystyle{elsarticle-num-names} 

\bibliography{ref}
\end{document}